\documentclass[12pt,draftclsnofoot,onecolumn]{IEEEtran}
\IEEEoverridecommandlockouts
\usepackage{cite}
\usepackage{amsmath,amssymb,amsfonts,amsthm}
\usepackage{lipsum}
\usepackage{mathtools}
\usepackage{cuted}
\usepackage{algorithm}
\usepackage{algorithmicx}
\usepackage{algpseudocode}
\usepackage{graphicx}
\usepackage{caption}
\usepackage{xcolor}
\usepackage{textcomp}
\usepackage{subfig}
\usepackage{tabu}
\usepackage{gensymb}

\def\BibTeX{{\rm B\kern-.05em{\sc i\kern-.025em b}\kern-.08em
    T\kern-.1667em\lower.7ex\hbox{E}\kern-.125emX}}
    
\begin{document}

{\Large \textbf{Notice:} This work has been submitted to the IEEE for possible publication. Copyright may be transferred without notice, after which this version may no longer be accessible.}
\clearpage

\title{Energy-Efficient Ultra-Dense Network with Deep Reinforcement Learning}
\author{
    \normalsize\IEEEauthorblockN{Hyungyu Ju\IEEEauthorrefmark{1}, Seungnyun Kim\IEEEauthorrefmark{1}, Youngjoon Kim\IEEEauthorrefmark{2}, and Byonghyo Shim\IEEEauthorrefmark{1}}
    \\
    \IEEEauthorblockA{\IEEEauthorrefmark{1}Seoul National University, Seoul, Republic of Korea\\
    }
    \IEEEauthorblockA{\IEEEauthorrefmark{2}Samsung Research, Seoul, Republic of Korea\\
    Email: \IEEEauthorrefmark{1}\{hkjoo, snkim, bshim\}@islab.snu.ac.kr, \IEEEauthorrefmark{2}yjoon100.kim@samsung.com}
\thanks{This work was supported by the National Research Foundation of Korea (NRF) grant funded by the Korea government (MSIT) (2020R1A2C2102198) and Samsung Research Funding \& Incubation Center for Future Technology of Samsung Electronics under Project Number (SRFC-IT1901-17).}
\thanks{Parts of this paper was presented at IEEE International
Workshop on Signal Processing Advanced in Wireless Communications (SPAWC), May 26-29, 2020~\cite{ju2020energy}.}}

\maketitle

\begin{abstract}
With the explosive growth in mobile data traffic, ultra-dense network (UDN) where a large number of small cells are densely deployed on top of macro cells has received a great deal of attention in recent years. While UDN offers a number of benefits, an upsurge of energy consumption in UDN due to the intensive deployment of small cells has now become a major bottleneck in achieving the primary goals viz., 100-fold increase in the throughput in 5G+ and 6G. In recent years, an approach to reduce the energy consumption of base stations (BSs) by selectively turning off the lightly-loaded BSs, referred to as the sleep mode technique, has been suggested. However, determining the appropriate active/sleep modes of BSs is a difficult task due to the huge computational overhead and inefficiency caused by the frequent BS mode conversion. An aim of this paper is to propose a deep reinforcement learning (DRL)-based approach to achieve a reduction of energy consumption in UDN. Key ingredient of the proposed scheme is to use decision selection network to reduce the size of action space. Numerical results show that the proposed scheme can significantly reduce the energy consumption of UDN while ensuring the rate requirement of network.
\end{abstract}
\newpage
\section{Introduction}

Densification of wireless network is a promising future direction to satisfy the explosive mobile data traffic demand.
In response to an increasing demand for data traffic along with the use of mmWave and terahertz bands, ultra-dense network (UDN) where a large number of small cells are densely deployed on top of macro cells has received special attention in recent years.
Since there are many small cells close to the mobile user, UDN reduces the path loss and also improves the quality of line-of-sight (LOS) transmission~\cite{kim2021energy}.
Furthermore, UDN expedites the reuse of spectrum per unit area dramatically, resulting in a significant improvement in the throughput of the wireless network~\cite{ji2016overview}.
However, energy consumption of UDN, caused by the exponential growth of a mobile data traffic, is a serious concern for the network operation~\cite{yang2020power}.
In fact, an upsurge of energy consumption is a heavy burden in the operational expense (OPEX) for the network operators, not to mention the increased carbon emission and the acceleration of global warming~\cite{ge20165g}.
Pursuing the balanced efficiency in energy consumption and throughput of UDN is, therefore, an important direction to ensure the sustainability of the next generation wireless communications~\cite{ge2015spatial}.

Among several factors contributing to the energy consumption of UDN, by far the dominant source is the base station (a.k.a eNB in 4G LTE and gNB in 5G NR). Indeed, it has been reported that more than half of the network energy is consumed at base station (BS).
Over the years, to reduce the energy consumption at the BS, a technique that deliberately turns off lightly loaded BSs, called the \textit{sleep mode energy saving technique}, has been proposed.
In~\cite{liu2015small}, a technique that randomly turns off the BSs has been proposed~\cite{han2011green, wu2015energy,Oh2016base}.
In~\cite{Oh2016base}, an approach that iteratively turns off the BSs with the lowest transmission power while guaranteeing the required mobile rate has been proposed.
However, in the UDN environment, energy saving might not be as dramatic as expected due to the exponential increase in computational complexity with the number of BSs.
In fact, active/sleep mode decision problem is a binary integer program (and hence NP-hard)~\cite{burer2012non} so that it is very difficult to find out the optimal active/sleep mode decision minimizing the energy consumption of UDN.
For this reason, to develop a technique that can effectively control the active/sleep mode of BSs in UDN while ensuring the rate requirement of mobile users is of great importance for the success of energy-efficient UDN.
\begin{figure}[t]
\centerline{\includegraphics[width = 150mm, height = 100mm]{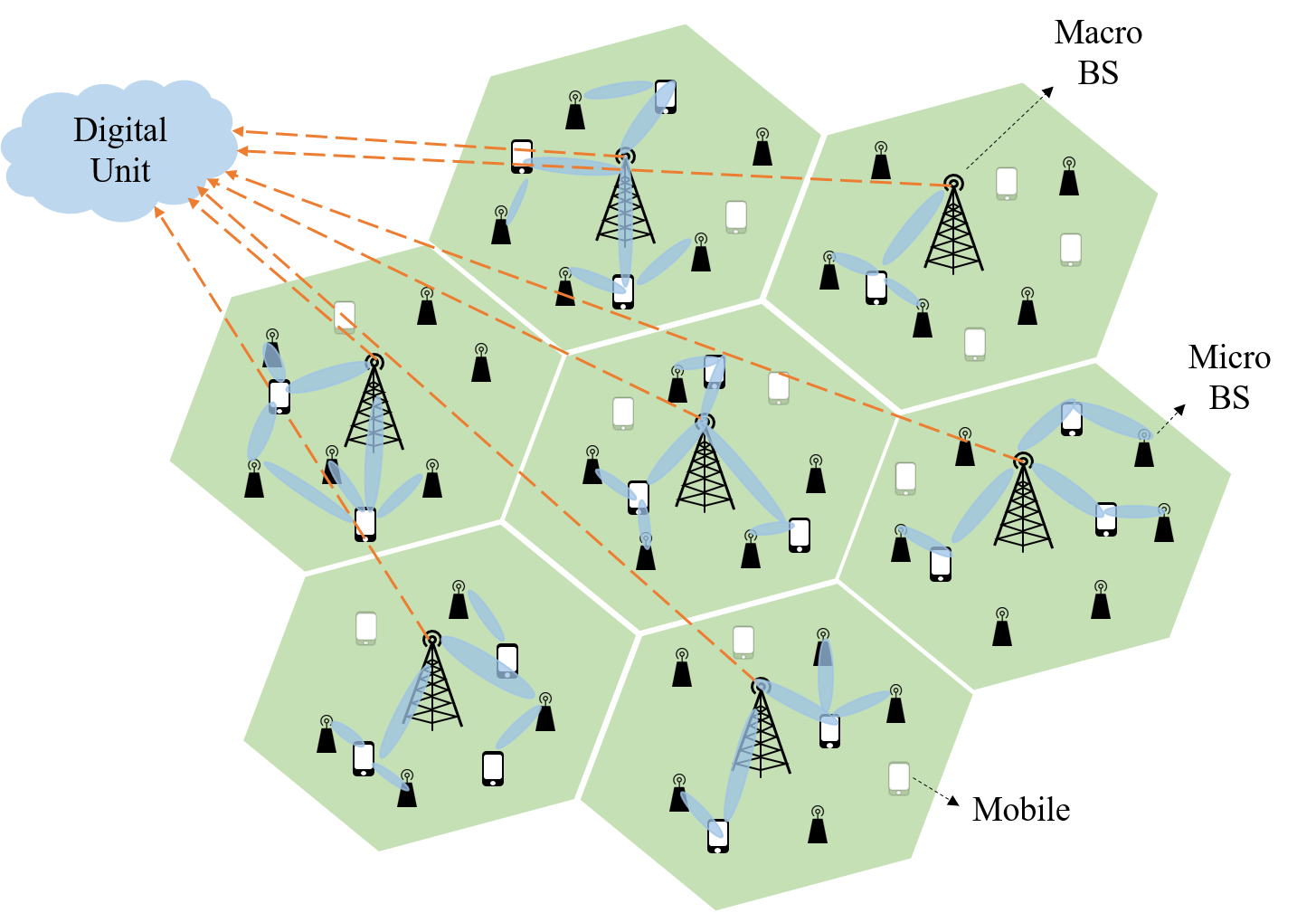}}
\captionsetup{justification=centering}
\caption{Illustration of UDN; The network consists of digital unit (DU) and various radio units (RU) including macro, micro and small BSs \label{fig:Hetnet}}
\end{figure}

An aim of this paper is to propose a deep learning (DL)-based approach to save the energy of UDN. 
While the conventional schemes aim to minimize the instantaneous energy consumption using the heuristic approach (e.g., proximity-based active/sleep mode control) and thus incur a significant waste of energy due to the frequent mode conversion, the proposed scheme, as referred to the \textit{deep reinforcement learning-based energy-efficient mode decision network} (DREEM-NET), achieves a reduction of the energy consumption over a long-term operational period.
In our study, we exploit the \textit{deep reinforcement learning} (DRL)~\cite{mnih2015human}, an efficient tool to solve the sequential decision-making problem, as a main engine.
In DRL, an agent, a component that makes a decision of what action to take, learns the optimal policy through the interactions with the environment. While the conventional RL cannot handle the large-scale control problems easily, DRL overcomes this limitation by replacing Q-table with the deep neural network (DNN).
In recent years, we have witnessed great success of DRL in various applications such as Go game~\cite{silver2017mastering}, natural language processing~\cite{young2018recent}, and resource management in computer systems and networks~\cite{mao2016rm}. DRL has also been applied to various wireless systems such as spectrum access~\cite{naparstek2017deep}, traffic scheduling~\cite{chinchali2018cellular}, and user association~\cite{zhao2018deep}\footnote{The active/sleep mode decision problem can be well modeled as a Markov decision process (MDP).}. 

Potential drawback of the DRL-based wireless systems is that an action (e.g., active/sleep mode decision) space size of the large-scale network is too large to handle the desired tasks.
In particular, when deciding the active/sleep mode of BSs in UDN, the number of possible choices increases exponentially with the number of BSs\footnote{For example, the number of possible decisions is $2^{20}\approx 10^{6}$ when we consider 20 cells in UDN.}. 
Due to the immense action space, during the training phase, a DRL agent is likely to explore undesirable actions (e.g., decision that cannot satisfy the mobile's rate requirement or decision that turns on too many BSs), resulting in a degradation in the energy efficiency and the slowdown of the convergence speed in training. 
In this work, to effectively reduce the size of action space, we introduce the notion of \textit{action elimination}~\cite{NIPS2018_7615}. 
Key idea of the action elimination is to identify undesirable actions (set of active/sleep modes violating the rate requirement or consuming large power) using DNN and then exclude them from the action space to avoid an exploration of such decisions. 
In doing so, we can improve the chance of experiencing the optimal or near optimal BS active/sleep decision minimizing the energy consumption of UDN. 

From the experimental results in the UDN environment, we observe that DREEM-NET reduces the energy consumption of UDN while guaranteeing the rate requirement of mobiles.
Specifically, DREEN-NET saves up to $20\%$ and $10\%$ of energy consumption against the full association scenario and the vanilla deep Q-network (DQN)-based method, respectively.
Even when compared to the conventional optimization-based energy saving technique, the energy consumption of DREEM-NET is about $18\%$ lower because it takes in account the transition power of BSs.

\textit{notations}: Lower and upper case symbols are used to denote vectors and matrices, respectively.
The superscript $\left(\cdot\right)^{\text{T}}$ and $\left(\cdot\right)^{\text{H}}$ denote transpose and Hermitian transpose, respectively. $\otimes$ and $\circ$ denote the Kronecker product and the Hadamard product, respectively. Also, $||x||$ denotes the Euclidean norm of a vector $x$ and $\mathbf{e}_m = \left[0, \cdots,1,\cdots,0\right]$ is an $M \times 1$ vector whose $m$-th element is one and otherwise zero.
\begin{table}[t]
\begin{center}
\caption{Summary of symbols}
\begin{tabu}{|c|c|c|c|}
\hline
\textbf{Notations} & \textbf{Values}  & \textbf{Notations}& \textbf{Values}\\ \hline
$R_{k,\text{min}}$ & Rate requirement of mobile $k$ & $\alpha_m$ & Mode decision of BS $m$ \\ \hline
$\eta_m$ & Amplifier efficiency of BS $m$& $\beta_{m,k}$ & Large-scale fading coefficient\\ \hline
$p_{m,k}$ & Power weight from the BS $m$ to the mobile $k$& $h_{m,k}$ & Downlink channel gain\\ \hline
$\rho^{\text{trans}}$ & Power consumed by mode transition & $P_{m,\text{max}}$ & Maximum transmission power of the BS $m$\\ \hline
$v_{\text{FT}}$ & Degree of infeasibility & $\gamma$ & Discount factor\\ \hline
$\mathcal{N}_R$ & Replay memory size & $\mathcal{A}$ & Total action space\\ \hline
$\mathcal{A}^F$ & Feasible action space & $\mathcal{A}^D$ & Desirable action space\\ \hline
\end{tabu}
\end{center}
\end{table}
\\
\section{Ultra-Dense Network System}
\subsection{System Model of UDN}
In this subsection, we discuss the system model of UDN.
We consider the downlink transmission where $M$ BSs equipped with a single antenna serve $K$ mobiles. 
In contrast to the cellular networks where a single BS serves the entire mobiles in one cell, a group of BSs cooperatively serves mobiles in UDN. 
Also, the mobiles move freely at a constant speed ($v \in$ Unif$\left[v_{\text{min}}, v_{\text{max}}\right]$) where $v_{\text{max}}$ and $v_{\text{min}}$ are the max/min speed of mobile.
A mobile changes its velocity when it reaches an edge of the service area.
The BSs (a.k.a radio unit or remote radio head (RRH)) are connected to a DU to share the channel state information (CSI) between the BSs and mobiles.
In order to indicate the active/sleep mode of BSs, we introduce the binary vector $\boldsymbol{\alpha}=\left[\alpha_{1},\cdots,\alpha_{M}\right]^{T}$ given by
\begin{align}
\alpha_{m}=
\begin{cases}
    1 & \text{if the BS}\,\, m\,\, \text{is in active mode} \\
    0 & \text{otherwise} \\
  \end{cases}.
\end{align}  
In our work, we consider the fading channel model where the downlink channel vector $h_{m,k}$ from the BS $m$ to the mobile $k$ is expressed as $h_{m,k}=\sqrt{\beta_{m,k}}g_{m,k}$
where $\beta_{m,k}$ is the large-scale fading coefficient and $g_{m,k}\sim\mathcal{CN}(0,1)$ is the small-scale fading coefficient. Note that the BS acquires the large-fading coefficients by averaging the channel magnitude extracted from the uplink pilot signals (e.g., SRS in 4G LTE). The transmit signal $x_{m}$ of the BS $m$ is 
\begin{align}\label{3.2.2}
x_{m}=\sum_{k=1}^{K}\sqrt{p_{m,k}}s_{k},
\end{align} 
where $s_{k}$ is the data symbol and $p_{m,k}\in\mathbb{R^{+}}\cup\{0\}$ is the power weight from the BS $m$ to the mobile $k$. Then, the received signal $y_{k}$ of the mobile $k$ is given by
\begin{align}\label{3.2.3}
y_{k}&=\sum_{m=1}^{M}h_{m,k}^{*}x_{m}+n_{k} \\
&=\sum_{m=1}^{M}\sqrt{p_{m,k}}h_{m,k}^{*}s_{k}+\sum_{j\neq k}^{K}\sum_{m=1}^{M}\sqrt{p_{m,j}}h_{m,k}^{*}s_{j}+n_{k},
\end{align}
where $n_{k}\sim\mathcal{CN}(0,\sigma_{k}^{2})$ is the additive Gaussian noise. The corresponding rate of mobile $k$ is given by
\begin{align}\label{3.2.5}
R_{k}&=\log_{2}\left(1+\frac{\sum_{m=1}^{M}\mathbb{E}\left[\lvert \sqrt{p_{m,k}}h_{m,k}^{*}\rvert^{2}\right]}{\sum_{j\neq k}^{K}\sum_{m=1}^{M}\mathbb{E}\left[\lvert \sqrt{p_{m,j}}h_{m,k}^{*}\rvert^{2}\right]+\sigma_{k}^{2}}\right) \\
&=\log_{2}\left(1+\frac{\sum_{m=1}^{M}p_{m,k}\mathbb{E}\left[\lvert h_{m,k}^{*}\rvert^{2}\right]}{\sum_{j\neq k}^{K}\sum_{m=1}^{M}p_{m,j}\mathbb{E}\left[\lvert h_{m,k}^{*}\rvert^{2}\right]+\sigma_{k}^{2}}\right) \\
&=\log_{2}\left(1+\frac{\sum_{m=1}^{M}p_{m,k}\beta_{m,k}}{\sum_{j\neq k}^{K}\sum_{m=1}^{M}p_{m,j}\beta_{m,k}+\sigma_{k}^{2}}\right).\label{3.2.5.3}
\end{align}
Note that the proposed scheme can be easily extended to the multiple-input multiple-output (MIMO) scenarios by replacing the power weight $p_{m,k}$ with the beamforming vector~\cite{ngo2017total}.
\
\subsection{Power Consumption Model and Problem Formulation in UDN}
The power consumption at the BS is divided into three parts: 
1) transmission power $P_{m}^{\text{tx}}$ consumed by the power amplifier and RF circuitry. 
2) active/sleep mode power $P_{m}^{\text{mode}}$ consumed by the signal processing, power supply, and air conditioning, and 3) transition power $P_{m}^{\text{trans}}$ consumed due to the mode transition (active to sleep mode or sleep to active mode). Combining these, the total power consumption of the BS $m$ at the time slot $t$ is $P_m^{(t)} = P_m^{\text{tx},(t)} + P_m^{\text{mode},(t)} +P_m^{\text{trans},(t)}$.

First, the transmission power of the BS $m$, consumed by power amplifier and RF circuitry, is expressed as
\begin{align}\label{3.2.5.4}
P_{m}^{\text{tx}}&=\frac{1}{\eta_m}\mathbb{E}\left[\lvert x_{m}\rvert^{2}\right] =\frac{1}{\eta_m}\sum_{k=1}^{K}p_{m,k},
\end{align}
where $\eta_m\in[0,1]$ is the power amplifier efficiency of the BS $m$. Depending on the type of BS, $40\sim50\%$ of the transmission power is used for transmission and the rest is wasted by heat~\cite{conte2012power}.
Second, the active/sleep mode power of the BS $m$, consumed by power supply and air conditioning, is given by 
\begin{align}\label{P_mode}
P_{m}^{\text{mode}}
&=\alpha_{m}P_{m}^{\text{on}}+(1-\alpha_{m})P_{m}^{\text{off}}.
\end{align}
where $P_{m}^{\text{on}}$ and $P_{m}^{\text{off}}$ are the power consumption of the BS $m$ in active mode and sleep mode, respectively.
Roughly speaking, about 35$\%$ of total power is consumed for this~\cite{conte2012power}.
Lastly, the mode transition power of the BS $m$, consumed by switching on and off mode of BS, is expressed as
\begin{align}\label{P_trans_1}
P_{m}^{\text{trans}}=\lvert\alpha_{m}-\alpha_{m}^{\text{prev}}\rvert\rho_{m}^{\text{trans}},
\end{align}
where $\alpha_{m}^{\text{prev}}$ is the active/sleep mode of BS $m$ in previous time slot and $\rho_{m}^{\text{trans}}$ is the power consumed by the mode transition of the BS $m$.
It has been shown that roughly 15$\%$ of total power is consumed for the mode transition~\cite{conte2012power}.

In summary, the total power consumption of the network $P_{\text{tot}}$ is given by
\begin{align}\label{P_total_1}
P_{\text{tot}}&=\sum_{m=1}^{M}\left(P_{m}^{\text{tx}}+P_{m}^{\text{mode}}+P_{m}^{\text{trans}}\right) \\
&=\sum_{m=1}^{M}\left(\frac{1}{\eta_m}\sum_{k=1}^{K}p_{m,k}+\alpha_{m}P_{m}^{\text{on}}+(1-\alpha_{m})P_{m}^{\text{off}}+\lvert\alpha_{m}-\alpha_{m}^{\text{prev}}\rvert\rho_{m}^{\text{trans}}\right).
\end{align}
Since many of the conventional BS sleep mode techniques focus on the reduction of the instantaneous power consumption (transmission power $P^{\text{tx}}$ and maintenance power $P^{\text{mode}}$), an energy consumption caused by the frequent mode transition is unavoidable. 

In order to jointly minimize the mode transition power and the instantaneous power, we pursue a reduction of the total power consumption of the entire BS during $T$ time slots:
\begin{align}
E_{total} = \sum_{t=1}^{T}\sum_{m=1}^{M}P_{\text{m}}^{(t)}.
\end{align}
The corresponding energy minimization problem is formulated as
\begin{subequations}
\begin{align}
\mathcal{P}_{1}:\underset{\lbrace\alpha_{m}^{(t)}\rbrace,\lbrace p_{m,k}^{(t)}\rbrace}{\text{min}}&E_{total}\\
\text{s.t.}\,\quad\,\,\,  &R_{k}^{(t)}\geq R_{k,\text{min}}^{(t)},\quad \forall k=1,\cdots,K\,,\,\forall t=1,\cdots,T\\
&P_{m}^{\text{tx},(t)}\leq P_{m,\text{max}},\quad \forall m=1,\cdots,M\,,\forall t=1,\cdots,T \\
&P_{m,k}^{(t)}\geq 0,\quad \forall m=1,\cdots,M \,,\forall k=1,\cdots,K ,\forall t=1,\cdots,T
\end{align}
\end{subequations}
where $R_{k,\text{min}}^{(t)}$ is the rate requirement of the mobile $k$ at time slot $t$ and $P_{m,\text{max}}$ is the maximum transmission power of the BS $m$. Since the active/sleep mode decisions $\lbrace\alpha_{m}^{(t)}\rbrace$ are binary integers, $\mathcal{P}_{1}$ is a mixed-integer programming that can be classified as a non-convex NP-hard problem. 
Therefore, to solve the problem using combinatoric approach is computationally prohibitive.
For example, if the number of coordinated BSs and time slots are 20 and 10, respectively, then one should search over humongous decision space (size is $2^{20 \cdot 10}\approx 10^{60}$) to find out the optimal active/sleep mode decision. Further, the analytic approach has a causality issue since the future-oriented active/sleep mode decisions require the channel information of future time slots.
\\
%

\section{Energy-Efficient Ultra-Dense Network Using Deep Reinforcement Learning}
The primary goal of this work is to find out the BS active/sleep modes minimizing the cumulative energy consumption. To achieve this goal, the proposed DREEM-NET exploits DRL framework in the BS active/sleep mode decision. 
DRL is a DL technique that learns the optimal policy for the sequential decision making problem through the interaction with the environment. 
Specifically, based on the input information (e.g., CSI, the required mobile rate), DNN in the DRL agent (i.e., DQN) learns the complicated input-output relationship between the current BS active/sleep decision and the cumulative energy consumption. A major hurdle in the DRL-based BS active/sleep mode decision framework is the action space that increases exponentially with the number of BSs. 
Since the DRL agent learns the policy by the trial and error, performance of DRL policy depends heavily on the exploration process of action space. 
In our case, due to the immense action space (e.g., $2^{10}\approx 1000$ active/sleep mode decisions when we consider $10$ cells), DRL agent needs to explore too many undesirable actions (e.g., active/sleep mode decision that cannot satisfy the mobile's rate requirement or decision that turns on unnecessary BSs).
Clearly, lack of useful training data lowers the sample efficiency\footnote{Sample efficiency indicates the amount of experience that an agent or algorithm needs to generate during training phase in order to achieve a certain level of performance.}, causing a slowdown of the training convergence and sub-optimal active/sleep mode control policy. 
\begin{figure}[t]
\centerline{\includegraphics[width=\columnwidth, height = 85mm]{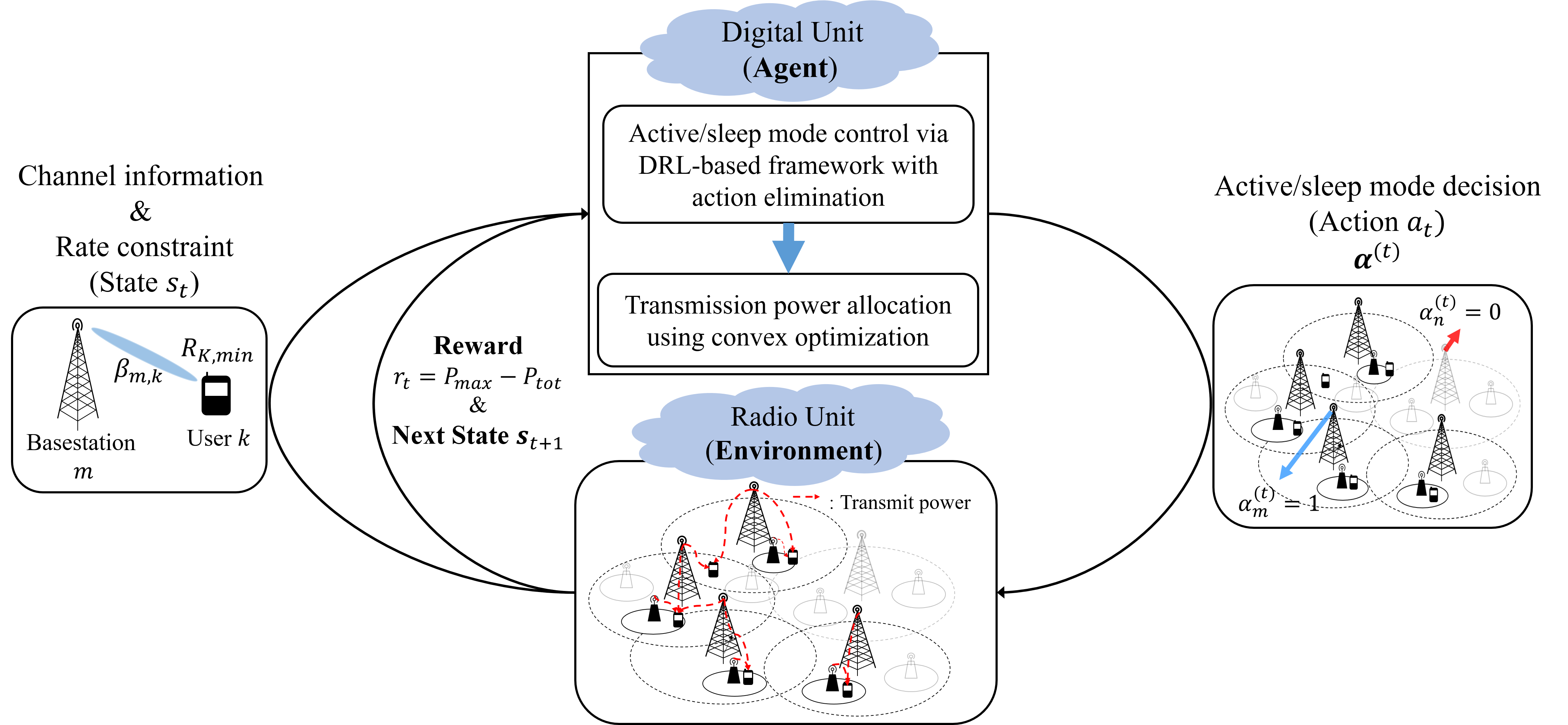}}
\captionsetup{justification=centering}
\caption{Basic structure of DRL-based energy-efficient UDN \label{fig:EntireSys}}
\end{figure}

To overcome this problem, we propose a \textit{decision selection network (DSN)} that identifies the undesirable active/sleep mode decisions and then eliminates them from the action space. 
Two key ingredients of the proposed DSN are 1) feasibility test to check whether the chosen active/sleep mode decision can satisfy the mobile's rate requirement and 2) energy consumption test to check whether the amount of energy consumed by the selected active/sleep mode decision exceeds the properly chosen threshold. 
Only the active/sleep modes that pass both tests are explored by the DRL agent. 
In doing so, we can dramatically reduce the action space and also improve the convergence speed. Since the DRL agent only explores the active/sleep mode decisions in desirable action space, the sample efficiency can be enhanced significantly, resulting in an improvement of the energy saving performance. 

Overall processes of the proposed DREEM-NET are summarized as follows: 1) the agent collects the CSI and the rate requirement through the backhaul link, 2) DSN filters out the undesirable active/sleep mode decisions and then the agent determines the active/sleep mode of BSs $\lbrace \alpha_{m}^{(t)}\rbrace$ from the input data, and 3) the agent allocates the transmit power $\lbrace p_{m,k}^{(t)}\rbrace$ for the BSs in active mode using the convex optimization.
4) Finally, the agent computes the reward based on the total power consumption of the network and then adjusts the active/sleep policy to maximize the cumulative reward (see Fig.~\ref{fig:EntireSys}).
\

\subsection{Basics of Reinforcement Learning}
In this subsection, we briefly explain basics of DRL. Reinforcement learning (RL) is a goal-oriented algorithm that learns how to achieve a goal using trials and errors.
Basic components of RL are agent, environment, state, action, and reward~\cite{sutton2018reinforcement}.
An agent learns the optimal policy for the sequential decision through the interaction with the environment. 
In the learning process of RL, an agent observes the current state  $s_t$, takes an action $a_t$, and then the environment returns the next state $s_{t+1}$ and the immediate reward $r_t$ to the agent as a feedback. A policy $\pi$ is a strategy that an agent uses to determine the action based on the current state. The goal of the agent is to find out the optimal policy $\pi^{*}$ maximizing the expected cumulative reward~\cite{sutton2018reinforcement}:
\begin{align}
\pi^{*} = \arg\max_{\pi}\,\,\mathbb{E}\left[\sum_{t=0}^{\infty} \gamma^{t} r_t \vert \pi\right],
\label{eqn.3.1}
\end{align}
where $\gamma$ is a discount factor ($0<\gamma<1$) to provide less weight to the future reward.

To find out the optimal policy $\pi^{*}$, the Q-value function $Q^{\pi}(s,a)$ that represents the expected cumulative reward obtained by following the policy $\pi$, is used:
\begin{align}
Q^{\pi}(s,a) = \mathbb{E}\left[\sum_{t=0}^{\infty} \gamma^{t} r_t \vert s_0 = s, a_0 = a\right] .
\label{eqn:opt2}
\end{align}
Since the Q-value function indicates the expected cumulative reward for taking action $a$ in state $s$, the optimal policy can be readily obtained by choosing the action maximizing $Q^{\pi}(s,a)$. To do so, the Q-value function should be available for all state-action pairs. 
To find out the optimal Q-function $Q^{*}(s,a)$, Bellman equation for $Q^{*}(s,a)$ is used~\cite{sutton2018reinforcement}:
\begin{align}
Q^{*}(s,a) = r(s,a) + \gamma\sum_{s'\in S} P^{a}_{ss'}\max_{a' \in A} Q^{*}(s',a'),
\label{eqn:opt3}
\end{align}
where $r(s,a)$ is the reward corresponding to the state-action pair ($s$, $a$) and $P^{a}_{ss'}$ is the transition probability defined as $P_{ss'}^{a}=P(s'\vert s, a)$. 

Finding out the optimal policy by directly solving the Bellman equation requires accurate knowledge of environment dynamics (i.e., reward function $r(s,a)$ and transition probability $P_{ss'}^{a}$).  
Unfortunately, in many practical scenarios, it is very difficult to acquire accurate $r(s,a)$ and $P_{ss'}^{a}$. Even though these quantities are acquired, considerable training overhead is required to solve the Bellman equation since we need to compute the Q-value for every state and action. 

To overcome the problem, Q-learning, a heuristic approach based on the trials and errors, has been proposed~\cite{sutton2018reinforcement}.
To be specific, $Q^{*}(s,a)$ can be approximated by replacing (\ref{eqn:opt3}) with the recursive update rule:
\begin{align}
Q_{t+1}(s,a) \leftarrow Q_t(s,a) + \kappa[r(s,a)+\gamma\max_{a' \in A} Q_t(s',a')],
\label{eqn:opt4}
\end{align}
where $\kappa$ is the learning rate. 
Key idea behind this update rule is to minimize the difference between the target Q-value (i.e., $r(s,a)+\gamma\max_{a' \in A} Q_t(s',a')$) and the current Q-value (i.e., $Q_t(s,a)$).

As mentioned, when the state space is large, it is very difficult to compute Q-values of all possible state-action pairs $(s,a)$. As a solution to the problem, DQN has been suggested~\cite{mnih2015human}. Main idea of DQN is to estimate Q-function using the DNN-based function approximator (i.e., $Q^{*}(s,a) \approx Q(s,a,w)$). The weight $w$ of DQN is updated to minimize the loss function $L(w)$ given by $L(w) = (Y^{dqn}_t - Q(s,a,w))^2$ where $Y^{dqn}_t = r(s,a)+\gamma\max_{a' \in A} Q(s',a',w)$.
\

\subsection{Energy-Efficient UDN Model}
In this subsection, we discuss the state space, action space, and reward function of the energy-efficient UDN. In the proposed DREEM-NET, UDN consisting of $M$ BSs and $K$ mobiles is considered as an environment and DU is used as a DRL agent (see Fig.~\ref{fig:EntireSys}).

\subsubsection{State Space} 
state contains essential information in the environment used for the policy learning. In the proposed DRL-framework, the state of the environment observed by the agent consists of several parts: the rate constraints of mobile $\boldsymbol{R}^{(t)}=[R_{1,min}^{(t)}, \cdots, R_{K, min}^{(t)}]^T$ at the time slot $t$, BS active/sleep mode decision at the previous time slot $\boldsymbol{\alpha}^{(t-1)}=[\alpha_{1}^{(t-1)}, \cdots, \alpha_{m}^{(t-1)}]^T$, and the large-scale fading coefficient matrix $\boldsymbol{H}^{(t)}$ reflecting on the path loss and the shadowing effect between BSs and mobiles at the time slot $t$. In particular, $\boldsymbol{H}^{(t)}$ is expressed as $M\times K$ matrix as
\begin{align}
\boldsymbol{H}^{(t)}=
  \begin{bmatrix}
    \beta_{1,1}^{(t)} & \dots  & \beta_{1,K}^{(t)}\\
    \vdots & \ddots & \vdots \\
    \beta_{M,1}^{(t)} & \dots  & \beta_{M,K}^{(t)}\\
  \end{bmatrix}
  .
\label{state_1}
\end{align}
In summary, the state can be expressed as
\begin{align}
s_t= [\boldsymbol{H}^{(t)}\ \boldsymbol{H}^{(t-1)}\ \boldsymbol{R}^{(t)}\ \boldsymbol{\alpha}^{(t-1)}]^T .
\end{align}

Note that both $\boldsymbol{H}^{(t)}$ and $\boldsymbol{H}^{(t-1)}$ are included in $s_{t}$ so that the DNN in DRL agent can extract the temporally correlated features in the channel such as angle of departure (AoD), delay spread, and path gain. By exploiting the extracted features between $\boldsymbol{H}^{(t)}$, $\boldsymbol{R}^{(t)}$, and $\boldsymbol{\alpha}^{(t-1)}$, the DRL agent learns the policy minimizing the instantaneous power and the mode transition power.

\subsubsection{Action Space} an action $a_t$ is defined as
\begin{align}
a_{t}=\boldsymbol{\alpha}^{(t)}=[\alpha_1^{(t)}\ \dots \ \alpha_M^{(t)}], \nonumber
\end{align}
where $\alpha_m^{(t)}=1$ (or $\alpha_m^{(t)}=0$) indicates that $m$-th BS is in active mode (or sleep mode). If we denote the set of possible actions as $A$, then the size of $A$ (i.e., the number of possible actions) is $2^{M}$. For example, if we control active/sleep modes of 20 BSs in UDN, then the size of $A$ is $2^{20}\approx 10^{6}$, which is clearly prohibitive. To deal with this so called \textit{curse of dimensionality}, we reduce the action space by excluding undesirable active/sleep mode decisions (we will say more in the next subsection).

\subsubsection{Reward} when the action of a time slot is decided, DU measures the power consumption of the network. Since the excessive power consumption should be penalized, we set the reward as
\begin{align*}
r_t &= P_{\text{max}} - P_{\text{tot}} \\ \nonumber
&= P_{\text{max}} - \sum_{m=1}^{M}(P_{m}^{\text{tx}}+P_{m}^{\text{mode}}+P_{m}^{\text{trans}}), \nonumber
\end{align*}
where $P_{\text{max}}$ is the total power consumption when all BSs are turned on.\footnote{However, if the active BSs cannot serve the mobiles, we impose a strong penalty (e.g., $r_t=-1000$) to the corresponding action.}
Since all three components of $P_{\text{tot}}$ are the function of active/sleep mode of BSs, the reward maximization problem is equivalent to the problem to find out the active/sleep mode minimizing 	$P_{\text{tot}}$.
\

\subsection{Importance of Efficient Exploration in DRL}
%

It is worth mentioning that when using DQN, the number of episodes required to find out accurate Q-values of all possible state-action pairs ($s$, $a$) scales linearly with the size of action space. 
In our case, due to the immense action space that increases exponentially with the number of BSs, it is very difficult to carry out sufficient number of episodes for the Q-value estimation. In the Atari game, for example, the number of possible actions is at most $18$ so that the sample efficiency is much higher than our UDN control problem~\cite{mnih2015human}. 
In our case, even though the DRL agent carries out numerous episodes, the chance of finding out the energy-efficient active/sleep mode decision policy is really tiny. 
The moral of the story is that an intelligent mechanism to properly control the action space is crucial for the success of our approach.
\\

\section{Action Space Reduction Via Decision Selection Network}

\begin{figure*}[t]
\centerline{\includegraphics[width = \linewidth, height = 90mm]{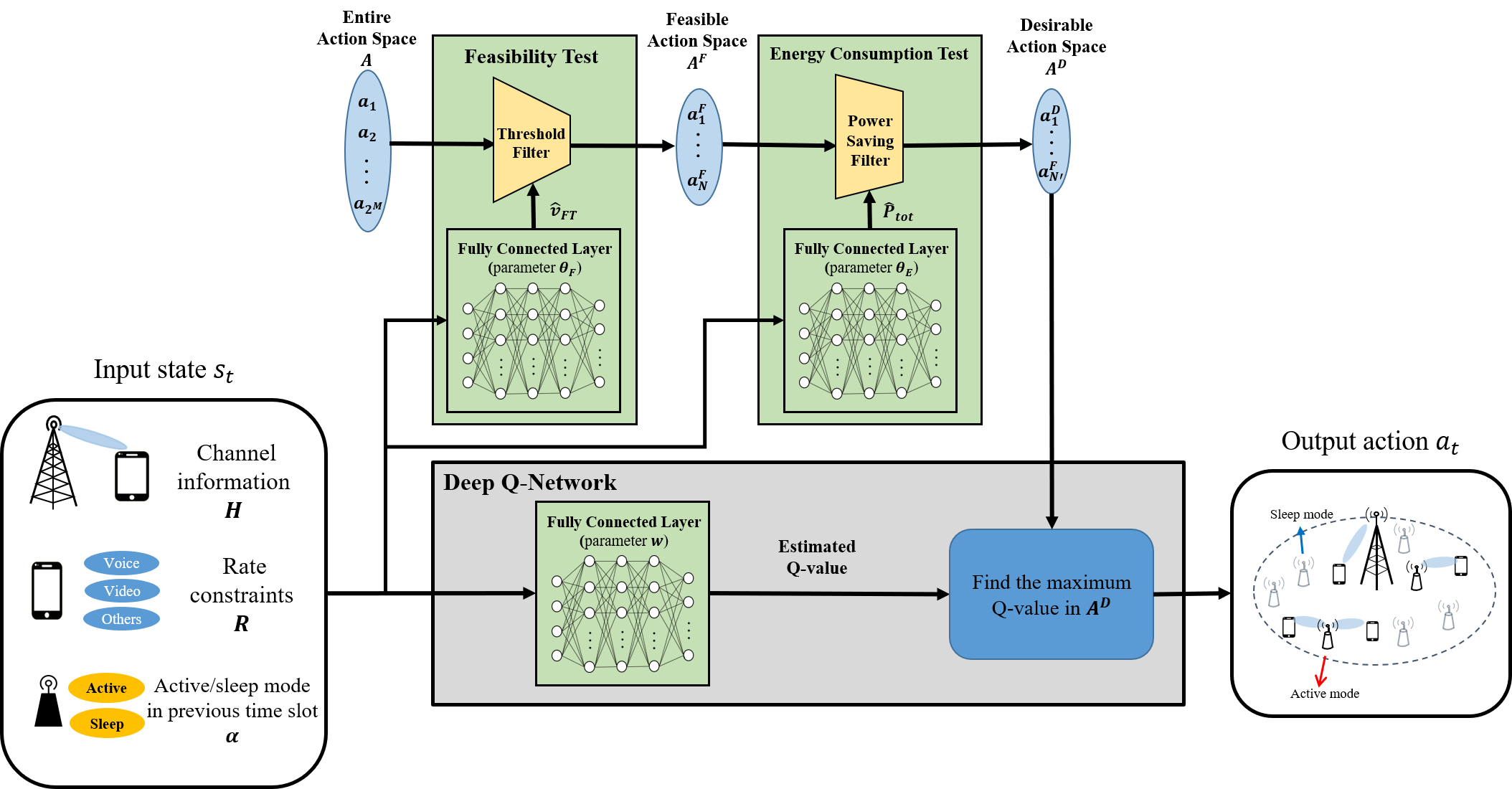}}
\captionsetup{justification=centering}
\caption{Deep Q-Network equipped with Decision Selection Network for Energy-Efficient UDN \label{fig:DRLSys1}}
\end{figure*}
\subsection{DSN architecture}
The main purpose of proposed DSN is to reduce the action space by removing undesirable actions that cannot satisfy the rate requirement of mobile or turns on too many unnecessary BSs. 
In order to identify these, we perform two tests: 1) \textit{feasibility test} to remove the infeasible active/sleep mode decision that cannot satisfy the rate requirement of mobile and 2) \textit{energy consumption test} to remove the redundant active/sleep mode decision that turns on too many unnecessary BSs. The detailed operations of DSN are as follows (see Fig.~\ref{fig:DRLSys1}).
\subsubsection{Feasibility Test}
In this step, we check whether the rate requirement of a mobile can be satisfied for each active/sleep mode decision $\boldsymbol{\alpha}$ and then exclude the infeasible decisions violating the mobile's rate constraints. Let $\mathcal{A}=\lbrace \boldsymbol{\alpha}_{1},\cdots,\boldsymbol{\alpha}_{2^{M}}\rbrace$ be the set of all possible active/sleep mode decisions. To determine whether $\boldsymbol{\alpha}$ is infeasible, we measure the degree of infeasibility $v_{\text{FT}}$, an achievable lower bound of maximum rate constraint violation:
\begin{align}
v_{\text{FT}}(s,\boldsymbol{\alpha})=\min_{\lbrace p_{m,k}\rbrace}\max_{k=1,\cdots,K}\,\big(R_{k,\text{min}}-R_{k}\big),\quad\forall \boldsymbol{\alpha}\in\mathcal{A}, \label{degreeof}
\end{align}
where $s$ is the DRL state. One can see that the rate constraint is satisfied (i.e., $R_{k,\text{min}}-R_{k}\leq 0$ for every $k$) if and only if $v_{\text{FT}}(s,\boldsymbol{\alpha})\leq 0$.
So, when $v_{\text{FT}}(s,\boldsymbol{\alpha})> 0$, we remove the infeasible decision $\boldsymbol{\alpha}$ .

In order to acquire $v_{\text{FT}}(s,\boldsymbol{\alpha})$, we need to solve the combinatoric optimization problem \eqref{degreeof}. This, basically, requires the large active/sleep mode decision space (i.e., $\lvert\mathcal{A}\rvert=2^{M}$) so that solving this problem for every $\boldsymbol{\alpha}$ is computationally infeasible. 
As a remedy, we exploit DNN that approximates $v_{\text{FT}}$ as a function of active sleep mode decision $\boldsymbol{\alpha}$ and DRL state $s$. That is,
\begin{equation}
\hat{v}_{\text{FT}}(s,\boldsymbol{\alpha})=f(s,\boldsymbol{\alpha}\,;\boldsymbol{\theta}_{F}),
\end{equation}
where $f$ is the input-output relationship and $\boldsymbol{\theta}_{\text{FT}}$ is the set of weights and biases. 
When the estimate of $v_{\text{FT}}$ exceeds a pre-defined threshold $\tau$, we eliminate the corresponding active/sleep mode decision from the action space $\mathcal{A}$. The obtained feasible action space $\mathcal{A}^{F}$ is
\begin{equation}
\mathcal{A}^{F}=\lbrace \boldsymbol{\alpha}\in\mathcal{A}\mid \hat{v}_{\text{FT}}(s,\boldsymbol{\alpha})\leq \tau\rbrace,
\end{equation}
where $\tau$ is a small positive value\footnote{Note that due to the estimation error between $v_{\text{FT}}$ and $\hat{v}_{\text{FT}}$, we set $\tau$ as a positive value. In doing so, we can prevent the feasible actions ($v_{\text{FT}}\leq 0)$ from being removed.} (e.g., $\tau=0.01$). 

\subsubsection{Energy Consumption Test}
In this step, we measure the total power consumption of network for each active/sleep mode decision $\boldsymbol{\alpha}$ and then exclude the decisions that consume excessively large power from $\mathcal{A}^{F}$. Recall that the total power consumption of network $P_{\text{tot}}$ is expressed as 
\begin{equation}
P_{\text{tot}}(s,\boldsymbol{\alpha})=P^{\text{tx}}(s,\boldsymbol{\alpha})+P^{\text{mode}}(s,\boldsymbol{\alpha})+P^{\text{trans}}(s,\boldsymbol{\alpha}).
\end{equation}
While $P^{\text{mode}}$ and $P^{\text{trans}}$ can be directly obtained from~(\ref{P_mode}) and~(\ref{P_trans_1}), we need to find $P^{\text{tx}}$ by solving the following transmission power allocation problem:
\begin{subequations}
\begin{align}
\mathcal{P}_{\text{tx}}:\,P^{\text{tx}}(s,\boldsymbol{\alpha})=\min_{\lbrace p_{m,k}\rbrace}\,&\sum_{m\in\mathcal{B}_{\text{on}}(\boldsymbol{\alpha})}^{M}P_{m}^{\text{tx}}\\
\text{s.t.}\,\,\,\,&R_{k}\geq R_{k,\text{min}},\quad \forall k=1,\cdots,K\\
&P_{m}^{\text{tx}}\leq P_{m,\text{max}},\quad \forall m=1,\cdots, M,
\end{align}
\end{subequations}
where $\mathcal{B}_{\text{on}}(\boldsymbol{\alpha})=\lbrace m\mid \alpha_{m}=1,\,m=1,\cdots,M\rbrace$ is the set of active BS of $\boldsymbol{\alpha}$. As mentioned, it is very difficult to solve $\mathcal{P}_{\text{tx}}$ for all $\boldsymbol{\alpha}\in\mathcal{A}^{F}$ in every time slot. Thus, similar to the feasibility test we discussed in the previous subsection, we exploit DNN to approximate the total power consumption as a function of $\boldsymbol{\alpha}$ and DRL state $s$. That is,
\begin{equation}
\hat{P}_{\text{tot}}(s, \boldsymbol{\alpha})=g(s,\boldsymbol{\alpha}\,;\boldsymbol{\theta}_{E}),
\end{equation}
where $g$ is the input-output relationship and $\boldsymbol{\theta}_{E}$ is the set of weights and biases. 
Once we obtain $\hat{P}_{\text{tot}}(s, \boldsymbol{\alpha})$ for each $\boldsymbol{\alpha} \in \mathcal{A}^F$, we eliminate the redundant active/sleep mode decision consuming excessive power and thus obtain the desirable action space $\mathcal{A}^{D}$:
\begin{equation}
\mathcal{A}^{D}=\left\lbrace \boldsymbol{\alpha}\in\mathcal{A}^{F}\mid \hat{P}_{\text{tot}}(s,\boldsymbol{\alpha})\leq P^{\text{threshold}}\right\rbrace.
\end{equation}
where $P^{\text{threshold}}$ is the power threshold for the energy consumption test (in our simulations, we set $P^{\text{threshold}}=64$).
\

\subsection{Training of DREEM-NET}
An integral part of the proposed DREEM-NET is the training process optimizing the network parameters $\boldsymbol{\theta}_F, \boldsymbol{\theta}_E,$ and $\boldsymbol{w}$. 
In the training phase, the network parameters are updated to minimize the loss functions of DSN ($L(\boldsymbol{\theta}_F)$ and $L(\boldsymbol{\theta}_E)$) and DQN loss function $L(w)$.
Since DNN in both tests of DSN are trained to minimize the error between predictions (i.e., $\hat{v}_{\text{FT}}$ and $\hat{P}_{\text{tot}}$) and actual values (i.e., $v_{\text{FT}}$ and ${P}_{\text{tot}}$), the loss function can be expressed as
\begin{align}\label{loss_ft}
L(\boldsymbol{\theta}_{F}) = (v_{\text{FT}}(s_t, a_t) - \hat{v}_{\text{FT}}(s_t,a_t,\boldsymbol{\theta}_{F}))^2   
\end{align} 
\begin{align}\label{loss_ec}
L(\boldsymbol{\theta}_E) = (P_{\text{tot}}(s_t, a_t) - \hat{P}_{\text{tot}}(s_t,a_t,\boldsymbol{\theta}_E))^2.
\end{align}
When the loss functions are differentiable, which is true in our case, we employ the stochastic gradient descent (SGD) method to update the parameters.
The update operations of SGD for feasibility test, energy consumption test, and DQN can be expressed as
\begin{align}
\boldsymbol{\theta}_{F}^{t+1}=\boldsymbol{\theta}_F^t - \kappa_F\nabla_{\boldsymbol{\theta}_F} L(\boldsymbol{\theta}_F)
\end{align}   
\begin{align}
\boldsymbol{\theta}_{E}^{t+1}=\boldsymbol{\theta}_E^t - \kappa_E\nabla_{\boldsymbol{\theta}_E} L(\boldsymbol{\theta}_E)
\end{align}
\begin{align}
\boldsymbol{w}_Q^{t+1}=\boldsymbol{w}_Q^t - \kappa_Q\nabla_{\boldsymbol{w}_Q} L(\boldsymbol{w}_Q)
\end{align}
where $\kappa_F, \kappa_E$, and $\kappa_Q$ are the learning rates of feasibility test, energy consumption test in DSN, and DQN, respectively. 

In order to compute the loss function in (\ref{loss_ft}) and (\ref{loss_ec}), we should have the actual value of $v_{FT}$ and  $P_{tot}$, which can be obtained by solving the following convex optimization problems. 
\begin{figure*}[t]
\centerline{\includegraphics[width = \linewidth, height = 100mm]{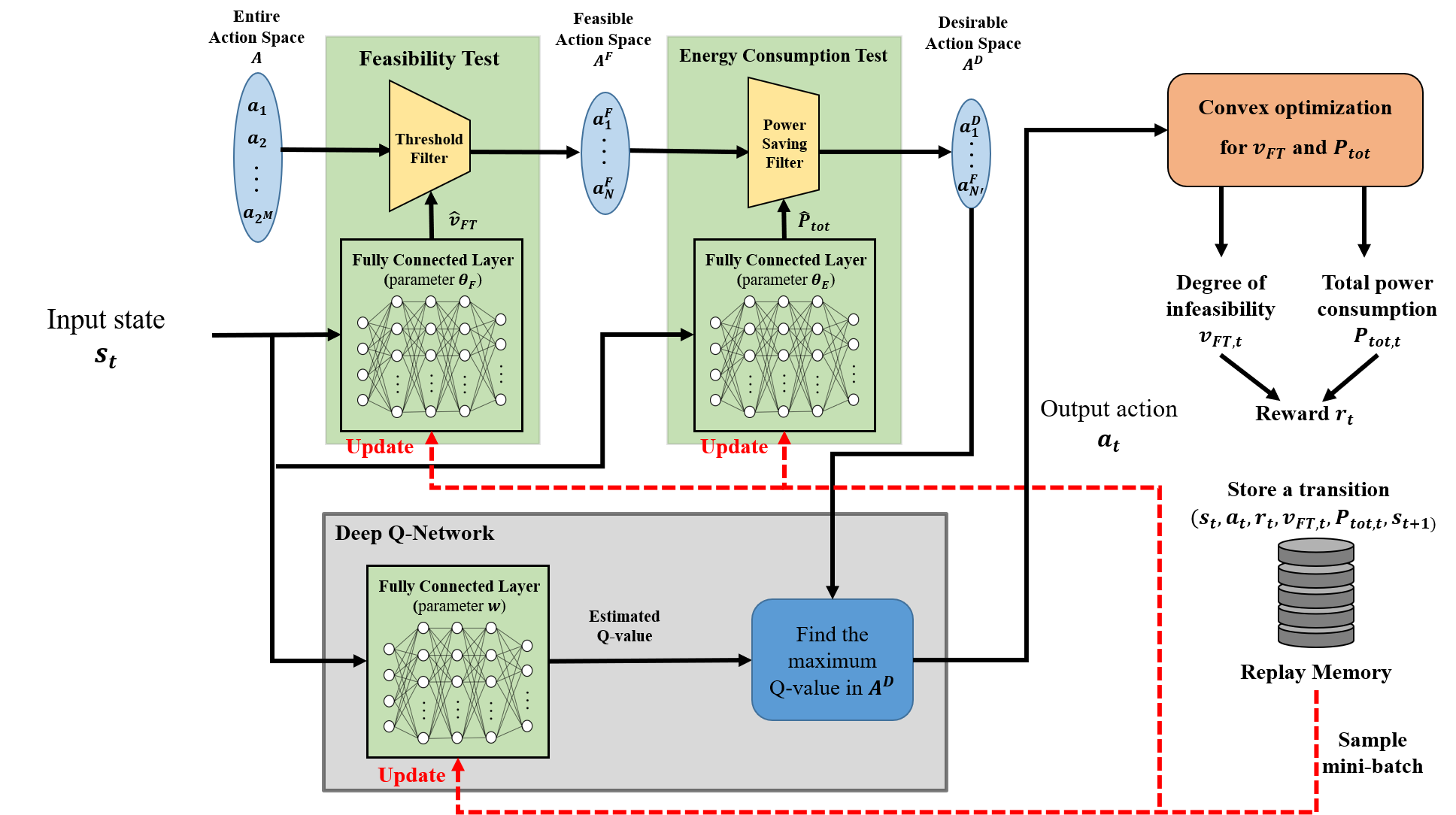}}
\captionsetup{justification=centering}
\caption{Training process of DREEM-NET : Red-dotted line indicates the training process \label{fig:DRLSys2}}
\end{figure*}
\subsubsection{Training Data Acquisition of Feasibility Test}

In order to obtain the degree of infeasibility, we need to solve the feasibility test problem $\mathcal{P}_{\text{FT}}$ for a given $\boldsymbol{\alpha}$, which is formulated as
\begin{subequations}
\begin{align}
\mathcal{P}_{\text{FT}}:v_{\text{FT}}^{*}=\underset{\lbrace p_{m,k}\rbrace}{\text{min}}\,\,\,&\underset{k}{\text{max}}\,\big(R_{k,\text{min}}-R_{k}\big) \label{3.3.1.2.1}\\
\text{s.t.}\,\,\,\,&P_{m}^{\text{tx}}\leq {\alpha}_{m}P_{m,\text{max}}, \quad \forall m=1,\cdots,M\,,\ \label{3.3.1.2.2}\\
&p_{m,k}\geq 0,\quad \forall m=1,\cdots,M,\,\ \forall k=1,\cdots,K. \label{3.3.1.2.3}
\end{align}
\end{subequations}
By concatenating the variables into a vector form (i.e., $\mathbf{p}_{k}=[p_{1,k},\cdots,p_{M,k}]^{\textrm{T}}$, $\mathbf{p}=[\mathbf{p}_{1}^{\textrm{T}},\cdots,\mathbf{p}_{K}^{\textrm{T}}]^{\textrm{T}}$), the rate constraint can be transformed to
\begin{align}
R_{k}\geq R_{k,\text{min}}&\iff \log_{2}\left(1+\frac{\sum_{m=1}^{M}p_{m,k}\beta_{m,k}}{\sum_{j\neq k}^{K}\sum_{m=1}^{M}p_{m,j}\beta_{m,k}+\sigma_{n}^{2}}\right)\geq R_{k,\text{min}} \label{3.3.1.3.1}\\
&\iff \boldsymbol{\beta}_{k}^{\textrm{T}}\Big(\mathbf{p}_{k}-\big(2^{R_{k,\text{min}}}-1\big)\sum_{j\neq k}^{K}\mathbf{p}_{j}\Big)\geq \sigma_{n}^{2}\big(2^{R_{k,\text{min}}}-1\big) \label{3.3.1.3.2}\\
&\iff \boldsymbol{\beta}_{k}^{\textrm{T}}(\mathbf{d}_{k}^{\textrm{T}}\otimes \mathbf{I}_{M})\mathbf{p}\geq \sigma_{n}^{2}\big(2^{R_{k,\text{min}}}-1\big), \label{3.3.1.3.3}
\end{align}
where $\boldsymbol{\beta}_{k}=[\beta_{1,k},\cdots,\beta_{M,k}]^{\textrm{T}}$ and $\mathbf{d}_{k}$ is a $K\times 1$ vector whose $k$-th element is one and others are $-(2^{R_{k,\text{min}}}-1)$. Thus, $\mathcal{P}_{\text{FT}}$ is reformulated as
\begin{subequations}
\begin{align}
\mathcal{P}_{\text{FT}}:v_{\text{FT}}^{*}=\underset{\mathbf{p}}{\text{min}}\,\,\,&\underset{k}{\text{max}}\,\big(\sigma_{n}^{2}\big(2^{R_{k,\text{min}}}-1\big)-\boldsymbol{\beta}_{k}^{\textrm{T}}(\mathbf{d}_{k}^{\textrm{T}}\otimes \mathbf{I}_{M})\mathbf{p}\big) \label{3.3.1.4.1}\\
\text{s.t.}\,\,\,\,&(\mathbf{1}_{K}^{\textrm{T}}\otimes\mathbf{e}_{m}^{\textrm{T}})\mathbf{p}\leq \hat{\alpha}_{m}\eta_{m}P_{m,\text{max}},\quad m\in\mathcal{B} \label{3.3.1.4.2}\\
&\mathbf{p}\succeq \mathbf{0}_{MK}, \label{3.3.1.4.3}
\end{align}
\end{subequations}
where \eqref{3.3.1.4.2} is a vector form expression of the power constraint. Let $v_{\text{FT}}=\underset{k}{\text{max}}\,\big(\sigma_{n}^{2}\big(2^{R_{k,\text{min}}}-1\big)-\boldsymbol{\beta}_{k}^{\textrm{T}}(\mathbf{d}_{k}^{\textrm{T}}\otimes \mathbf{I}_{M})\mathbf{p}\big)$, then we have
\begin{subequations}
\begin{align}
\mathcal{P}_{\text{FT}}:v_{\text{FT}}^{*}=\underset{\mathbf{p},v_{\text{FT}}}{\text{min}}\,\,\,&v_{\text{FT}} \label{v_FT}\\
\text{s.t.}\,\,\,\,&\sigma_{n}^{2}\big(2^{R_{k,\text{min}}}-1\big)-\boldsymbol{\beta}_{k}^{\textrm{T}}(\mathbf{d}_{k}^{\textrm{T}}\otimes \mathbf{I}_{M})\mathbf{p}\leq v_{\text{FT}},\quad \forall k=1,\cdots,K \label{3.3.1.5.2}\\
&(\mathbf{1}_{K}^{\textrm{T}}\otimes\mathbf{e}_{m}^{\textrm{T}})\mathbf{p}\leq {\alpha}_{m}\eta_{m}P_{m,\text{max}},\quad \forall m=1,\cdots,M \label{3.3.1.5.3}\\
&\mathbf{p}\succeq \mathbf{0}_{MK}. \label{3.3.1.5.4}
\end{align}
\end{subequations}
Since the objective function and constraints are all linear functions of $\mathbf{p}$, $\mathcal{P}_{\text{FT}}$ is a linear programming (LP) problem that can be solved by the convex optimization tool (e.g., CVX).
\subsubsection{Training Data Acquisition of Energy Consumption Test}

In order to obtain the total power consumption $P_{\text{tot}}$ for given $\boldsymbol{\alpha}$, we solve the transmission power allocation problem, which is formulated as
\begin{subequations}
\begin{align}\label{7}
\mathcal{P}_{\text{tx}}:\,\underset{\lbrace p_{m,k}\rbrace}{\text{min}}\,\,\, &\sum_{m\in\mathcal{B}_{on}}P_{m}^{\text{tx}}\\
\text{s.t.}\quad &R_{k}\geq R_{k,\text{min}},\quad\forall k=1,\cdots,K \label{7.1.1}\\
&P_{m}^{\text{tx}}\leq P_{m,\text{max}},\quad\forall m=1,\cdots,M .\label{7.1.2}
\end{align}
\end{subequations}
Using the rate expression in \eqref{3.2.5.3} and the power consumption model in \eqref{3.2.5.4}, $\mathcal{P}_{\text{tx}}$ can be re-expressed as
\begin{subequations}
\begin{align}\label{P_trans}
\mathcal{P}_{\text{tx}}:\underset{\lbrace p_{m,k}\rbrace}{\text{min}}&\sum_{m\in\mathcal{B}_{on}}\frac{1}{\eta_m}\sum_{k=1}^{K}p_{m,k}\\ 
\text{s.t.}\,\,\, &\!\sum_{m\in\mathcal{B}_{on}}\beta_{m,k}p_{m,k}\!-\!\left(2^{R_{k,\text{min}}}\!-\!1\right)\!\sum_{j\neq k}^{K}\sum_{m\in\mathcal{B}_{on}}\beta_{m,k}p_{m,j}\nonumber\\
&\geq \sigma_{n}^{2}\left(2^{R_{k,\text{min}}}-1\right),\,\forall k=1,\cdots,K  \\
&\alpha_{m}\sum_{k=1}^{K}p_{m,k}\leq P_{m,\text{max}},\quad \forall m=1,\cdots,M  \label{2.2.2}.
\end{align}
\end{subequations}
Similar to the optimization problem in the feasibility test, the modified problem $\mathcal{P}_{\text{tx}}$ is an LP and thus can be solved by the convex optimization tool. Using the obtained $\lbrace p_{m,k} \rbrace$ and $\boldsymbol{\alpha}$, we can calculate $P_{\text{tot}}=P^{\text{tx}}+P^{\text{mode}}+P^{\text{trans}}$ (see~(\ref{P_total_1})).

After obtaining the degree of infeasibility $v_{\text{FT},t}$ in \eqref{v_FT} and the total power consumption $P_{\text{tot},t}$ in \eqref{P_trans}, the agent receives the immediate reward $r_t$ computed by the environment and the next state $s_{t+1}$ from the UDN environment.
In each time slot, the transition tuple $(s_t, a_t, r_t, v_{FT,t}, P_{\text{tot},t}, s_{t+1})$ observed by the agent is stored to the replay memory.
As shown in Algorithm 1, in each iteration of the training phase, a mini-batch data is randomly sampled from the replay memory and then the weights of DQN and DSN are updated in a direction to minimize the loss value in $L(w)$, $L(\boldsymbol{\theta}_F)$ and $L(\boldsymbol{\theta}_E)$. 

\subsection{Computational Complexity of DREEM-NET}
In this subsection, we analyze the computational complexities of DREEM-NET and convetional schemes including sequential on/off method~\cite{Oh2016base} and mixed-integer linear programming (MILP)-based on/off method. Since the convex optimization tool for the transmission power allocation among active BSs is used to all techniques ($\mathcal{C}_{\text{TX}}=M^3 K^3\log(\frac{1}{\epsilon}$) where $\epsilon$ is the CVX solver tolerance)~\cite{bubeck2014convex}, we focus on the complexity analysis of each technique during the active/sleep mode decision process.

We first analyze the complexity of DRL framework in DREEM-NET.
Initially, in the fully-connected layer for the feasibility test, the input state vector $s_t \in \mathbb{R}^{\left(2MK+M+K\right)\times 1}$ is multiplied by the weight $\boldsymbol{\theta}^{w}_F \in \mathbb{R}^{\omega \times \left(2MK+K+M\right)}$ and then the bias $\boldsymbol{\theta}^{b}_F \in \mathbb{R}^{\omega \times 1}$ is added where $\omega$ is the width of hidden layer.
The complexity of the input layer for the feasibility test $\mathcal{C}_ {F}$ is
\begin{align}
\mathcal{C}_{F,\text{in}} =(2 \left(2MK+M+K\right) -1)\omega + \omega = (4MK+2M+2K)\omega
\end{align}
Next, the complexity of the hidden layer (4 layers) and the output layer for the feasibility test $\mathcal{C}_{F,\text{hidden}}$ and $\mathcal{C}_{F,\text{out}}$, respectively, are
\begin{align}
\mathcal{C}_{F,\text{hidden}} =4((2\omega -1)\omega + \omega) = 8\omega^2
\end{align}
\begin{align}
\mathcal{C}_{F,\text{out}} =(2\omega -1)\vert\mathcal{A}\vert + \vert\mathcal{A}\vert = 2\omega\vert\mathcal{A}\vert
\end{align}
Thus, considering the threshold filter ($\vert\mathcal{A}\vert$ flops), the complexity of the feasibility test is
\begin{align}
\mathcal{C}_{F} =(4MK+2M+2K+8\omega+2\vert\mathcal{A}\vert)\omega+\vert\mathcal{A}\vert.
\end{align}
In the similar way, the complexity of energy consumption test and DQN can be expressed as 
\begin{align}
\mathcal{C}_{E} =(4MK+2M+2K+8\omega+2\vert\mathcal{A^F}\vert)\omega+\vert\mathcal{A^F}\vert
\end{align}
\begin{align}
\mathcal{C}_{\text{DQN}} =(4MK+2M+2K+8\omega+2\vert\mathcal{A^D}\vert)\omega+\vert\mathcal{A^D}\vert
\end{align}
Thus, the complexity of DREEM-NET is summarized as
\begin{align*}
\mathcal{C}_{\text{DREEM-NET}} &=  \mathcal{C}_{F} +\mathcal{C}_{E} +\mathcal{C}_{\text{DQN}} + \mathcal{C}_{\text{TX}} \\
&=(4MK+2M+2K+8\omega+2)3\omega + (\vert\mathcal{A}\vert+\vert\mathcal{A^F}\vert+\vert\mathcal{A^D}\vert)(2\omega+1)+(M^3 K^3\log(\frac{1}{\epsilon}))
\end{align*}

We next analyze the complexities of sequential on/off method and MILP-based method. From the literature~\cite{Oh2016base}, sequential on/off method turns on all BSs initially and then removes the BS having the minimum impact (i.e., minimum transmission power in our work) on the energy consumption one after another until it reaches to the point where the mobile's rate requirement is violated.
Then, the complexity of sequential on/off method is
\begin{align*}
\mathcal{C}_{\text{sequential}}&=\left(\sum_{i=0}^{\mathcal{N}_{\text{off}}}(M-i)\right)+(\mathcal{N}_{\text{off}}+2) \times \mathcal{C}_{\text{TX}} \\
&= \left(M-\frac{\mathcal{N}_{\text{off}}}{2}\right)\mathcal{N}_{\text{off}}+(\mathcal{N}_{\text{off}}+2)(M^3 K^3\log(\frac{1}{\epsilon})),
\end{align*}
where $\mathcal{N}_{\text{off}}$ is the number of sleep BSs.

MILP-based method uses CVX gurobi solver based on branch-and-bound algorithm~\cite{grant2009cvx}. The complexity of MILP-based method is as follows,
\begin{align*}
\mathcal{C}_{\text{MILP}}=2^{M \cdot K}(M^3 K^3\log(\frac{1}{\epsilon})).
\end{align*}

\begin{algorithm}
\begin{algorithmic}[1]
\item[\textbf{Input:}] CSI $\lbrace H^{(t)}\rbrace$, required mobile rate $\lbrace R_{\text{min}}^{(t)}\rbrace$, active/sleep mode at previous time slot $\boldsymbol{\alpha}^{(t-1)}$, DNN-based DSN $f$ and $g$ (weight $\boldsymbol{\theta}_F$ and $\boldsymbol{\theta}_E$), DRL-based active/sleep mode decision network $h$ (weight $\boldsymbol{w}
$), learning rate $\kappa_D$, $\kappa_w$, number of time slots $T$
\item[\textbf{Initialization:}] $t=0$, $\boldsymbol{\theta}_{t}=\boldsymbol{\theta}_{ini}$ 
\item[\textbf{Iteration:}] 
\While{$\boldsymbol{\theta}_{t}$ does not converge}
\For{$t=1,\cdots, T$}
\State State $s_{t}=[H^{(t)}, H^{(t-1)},\,R_{\text{min}}^{(t)},\,\boldsymbol{\alpha}^{(t-1)}]$
\State Exclude infeasible mode decision through the feasibility test using $f$
\State Exclude redundant mode decision through the energy consumption test using $g$  
\State Obtain $\boldsymbol{\alpha}^{(t)}$ by DQN based on reduced action space $\mathcal{A}^{D}$
\State Allocate the transmission power for the set of active BS $\mathcal{B}_{on}$
\State Solve $\mathcal{P}_{\text{FT}}$ for $\boldsymbol{\alpha}^{(t)}$ to obtain the degree of infeasibility ${v}_{\text{FT},t}$
\State Solve $\mathcal{P'}_{\text{tx}}$ to obtain total power consumption $P_{\text{tot},t}$ 
\State Compute reward $r_t$ and observe the next state $s_{t+1}$
\State Store the transition $(s_t, a_t, r_t, {v}_{\text{FT},t}, P_{\text{tot},t}, s_{t+1})$ into replay memory $\mathcal{R}$
\EndFor
\State Randomly sample a mini-batch of the transition $(s_i, a_i, r_i, {v}_{\text{FT},i}, P_{\text{tot},i}, s_{i+1})$ with a size 
\Statex  \quad \,\  of $\mathcal{N}_{\mathcal{R}}$
\State Compute $\nabla_{\boldsymbol{\theta}_F} L(\boldsymbol{\theta}_F)=\nabla_{\boldsymbol{\theta}_F}\sum_{i}(v_{\text{FT},i}(s_i, a_i) - \hat{v}_{\text{FT},i}(s_i,a_i,\boldsymbol{\theta}_F))^2$
\State Compute $\nabla_{\boldsymbol{\theta}_E} L(\boldsymbol{\theta}_E)=\nabla_{\boldsymbol{\theta}_E}\sum_{i}(P_{\text{tot},i}(s_i, a_i) - \hat{P}_{\text{tot},i}(s_i,a_i,\boldsymbol{\theta}_E))^2$
\State Compute $\nabla_{\boldsymbol{w}} L(w)=\nabla_{\boldsymbol{w}}\sum_{i}(r(s_i,a_i)+\gamma\max_{a_{i+1}} Q(s_{i+1},a_{i+1},\boldsymbol{w}) - Q(s_{i+1},a_{i+1},\boldsymbol{w}))^2$
\State ${\boldsymbol{\theta}}_{t+1}={\boldsymbol{\theta}}_{t}-\kappa_D\nabla_{\boldsymbol{\theta}} L(\boldsymbol{\theta})$
\State $\boldsymbol{w}_{t+1}=\boldsymbol{w}_{t}-\kappa_w\nabla_{\boldsymbol{w}} L(\boldsymbol{w})$
\State $t=t+1$
\EndWhile
\captionof{algorithm}{Training process of DREEM-NET}
\end{algorithmic}
\end{algorithm}

\section{Simulation Results}
\subsection{Simulation Setup}

In this section, we describe numerical results to evaluate the performance of DREEM-NET.
In our simulations, we consider the UDN scenario where $M$ small BSs simultaneously serve $K$ mobiles. The small cells are uniformly distributed in a square service area of $D \times D$ $\text{km}^2$ for UDN configuration~\cite{kamel2016ultra} and the mobiles move freely at a constant speed ($v \in$ Unif$[v_{\text{min}}, v_{\text{max}}]$ where $v_{\text{max}}$ and $v_{\text{min}}$ are the max/min speed of mobile.
Also, we use a full-buffer traffic model which is suitable for the situation where the number of mobiles in the coverage of macro BS is constant~\cite{ameigeiras2012traffic}.
For the fading channel model, we use the small-scale fading coefficient $g_{m,k}$ generated from the complex Gaussian distribution (i.e., $g_{m,k} \sim \mathcal{CN}(0,1)$) and the large-scale fading coefficient $\beta_{m,k}$ generated based on Hata-COST231 model~\cite{singh2012comparison}, which is expressed as 
\begin{align*}
\beta_{m,k} &= \text{PL}_{m,k}\cdot 10^{\frac{z_{m,k}\sigma_{sh}}{10}}
\end{align*}
where $\text{PL}_{m,k}$ is the path loss and $10^{\frac{z_{m,k}\sigma_{sh}}{10}}$ is the shadow fading ($z_{m,k}\sim \mathcal{N}(0,1)$).
Specifically, $\text{PL}_{m,k}$ is given by
\begin{align}
\text{PL}_{m,k}=
\begin{cases}
-L-35\log_{10}(d_{m,k}) & \text{if }d_{m,k}>d_{1}\\
-L-15\log_{10}(d_{1})-20\log_{10}(d_{m,k}) & \text{if }d_{0}<d_{m,k}\leq d_{1}\\
-L-15\log_{10}(d_{1})-20\log_{10}(d_{0}) & \text{if }d_{m,k}\leq d_{0}
\end{cases}
\end{align}
where $d_{m,k}$ is the distance between the BS $m$ and the mobile $k$ and
\begin{align*}
L &= 46.3-33.9\log_{10}f-13.82\log_{10}h_b -(1.1\log_{10}f-0.7)h_u-(1.56\log_{10}f-0.8)
\end{align*}
where $f$ is the carrier frequency, $h_B$ and $h_U$ are the heights of BS and mobile, respectively.
\begin{table}[t]
\begin{center}
\caption{Simulation parameters}
\begin{tabular}{c|c|c|c}
\hline
\textbf{Parameters} & \textbf{Values}  & \textbf{Parameters} & \textbf{Values}\\ \hline
Carrier frequency ($f$) & $2\,\text{GHz}$ & Number of time slots ($T$) & $50$ \\ \hline
BS height ($h_{B}$) & $15\,\text{m}$ & Maximum speed of mobile ($v_{\text{max}}$) & $6\,\text{m/s}$\\ \hline
Mobile height ($h_{U}$) & $1.65\,\text{m}$ & Amplifier efficiency ($\eta$) & $0.25$\\ \hline
Service area radius ($D$) & $200\,\text{m}$ & Active mode BS power ($P^{\text{on}}$) & $6.8\,\text{W}$ \\ \hline
Path loss variable ($d_{0}$) & $10\,\text{m}$ & Sleep mode BS power ($P^{\text{off}}$) & $4.3\,\text{W}$\\ \hline
Path loss variable ($d_{1}$) & $50\,\text{m}$ & Maximum transmission power ($P_{\text{max}}$) & $1\,\text{W}$\\ \hline
Shadow fading deviation ($\sigma_{sh}$) & $3\,\text{dB}$ & Mode transition power ($\rho^{\text{trans}}$) & $3\,\text{W}$ \\ \hline
Number of small cells ($M$) & $10$ & Gamma ($\gamma$) & $0.9$\\ \hline
Number of mobiles ($K$) & $4$ & Mini-batch size & 256\\ \hline
Noise power ($\sigma_{n}^{2}$) & $-174\,\text{dBm}/\text{Hz}$ & Replay memory size ($\mathcal{N}_{\mathcal{R}}$)& 20000\\ \hline
\end{tabular}
\end{center}
\end{table}
\\
Considering the transition latency (deactivation + reactivation latency) of BS~\cite{debaillie2015flexible}, we set the time interval of power measurement and the active/sleep mode decision as the coherence time of large-scale fading coefficient ($T_{\beta}=1.53s$). Note that the instantaneous power consumption is the power consumption consumed by the network during a single time slot whereas the total power consumption is the energy consumed by the network during $L$ consecutive time slots. In our simulations, we set $L$ to 50 so the time duration is about 75s.
As discussed, DREEM-NET consists of DSN and DQN, each of which consists of 6 fully connected layers (width of a hidden layer is set to 512).
In the network parameter training, we use Adam optimizer, a well-known optimization tool to guarantee the robustness of learning process. The simulation parameters are summarized in Table II.

We compare the proposed DREEM-NET with four baseline cell power control techniques: 1) \textit{full association} method where all BSs are in active mode, 2) \textit{sequential on/off} method that turns on all BSs initially and then removes the BS having the minimum impact on the energy consumption one after another until it reaches to the point where the mobile's rate requirement is violated~\cite{Oh2016base}, 3) \textit{MILP-based on/off} method where the BS active/sleep mode and the corresponding transmission power are optimized simultaneously in each time slot, and 4) vanilla DQN-based method where the active/sleep mode of BSs is determined by using the original DQN~\cite{xu2017deep}. For the transmission power allocation among active BSs, the convex optimization technique is used for all techniques under investigation.


\subsection{Simulation Result}
%

\begin{figure}[t]
\begin{minipage}[t]{0.48\linewidth}
    \includegraphics[width=\linewidth, height =70mm]{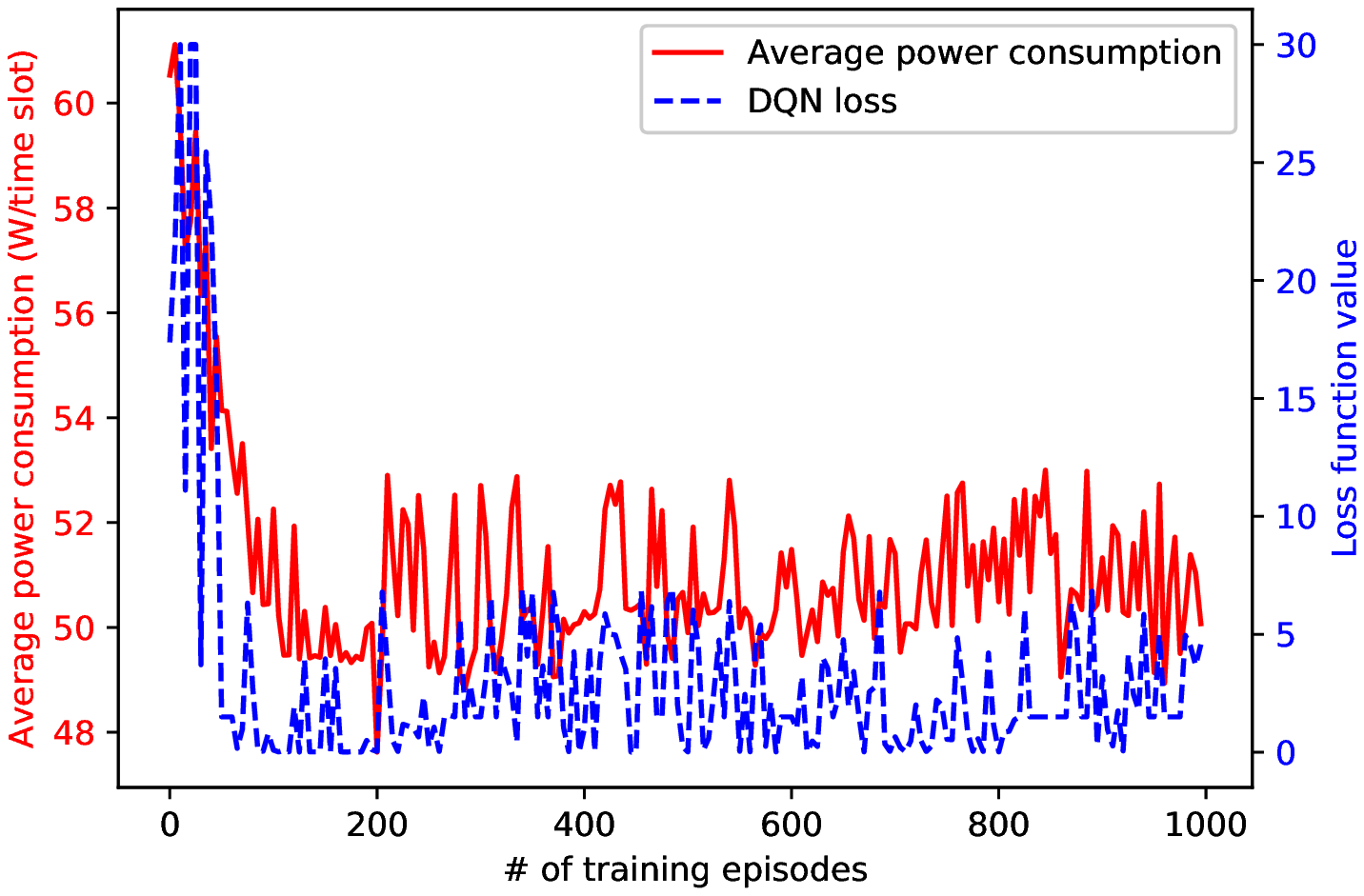}
    \caption{Average power consumption and DQN loss as functions of the number of training episodes}
    \label{fig:fig5}
\end{minipage}  
    \hfill%
\begin{minipage}[t]{0.48\linewidth}
    \includegraphics[width=\linewidth, height =70mm]{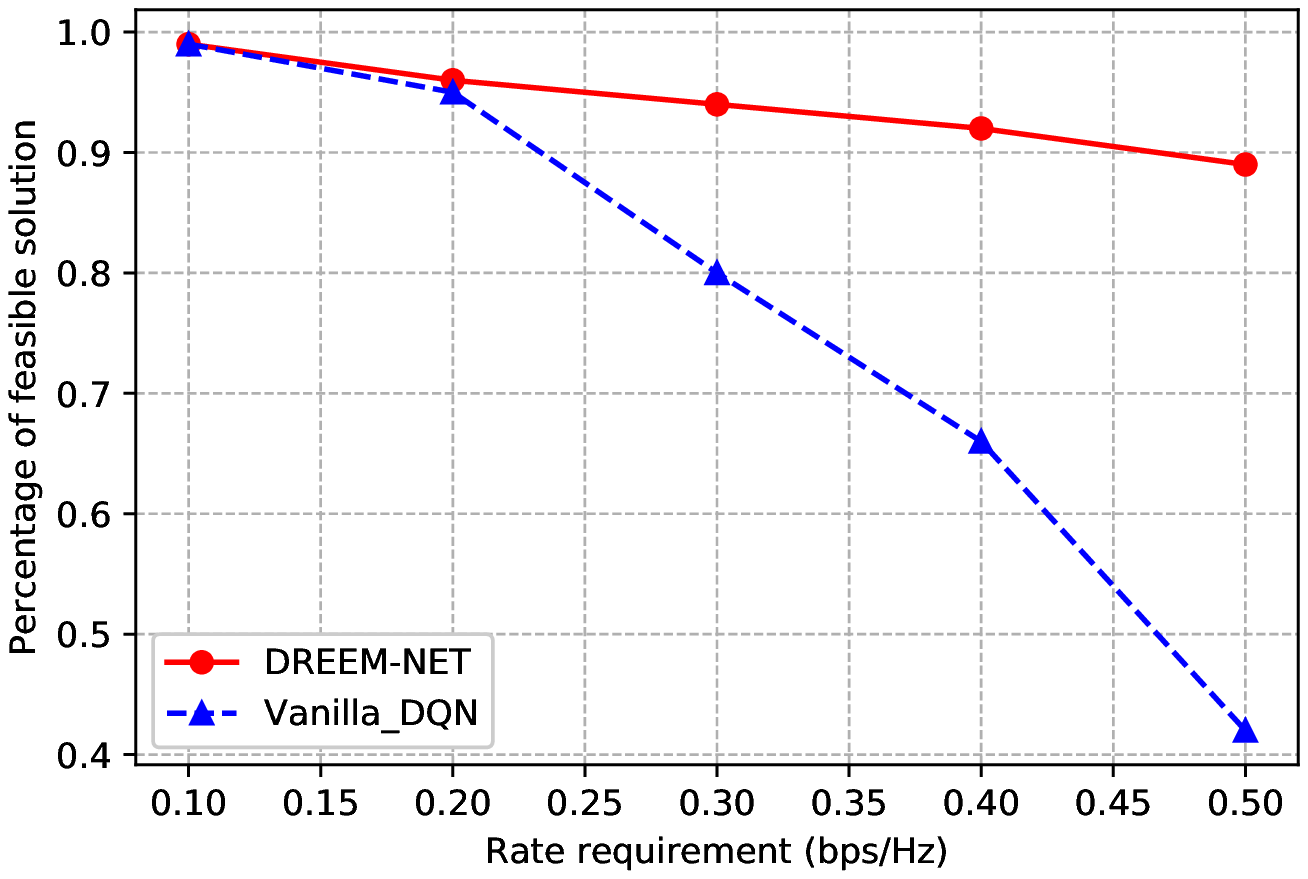}
    \caption{percentage of feasible solution in the action space ($M=10$, $K=4$, and $\text{SNR}=10\,\text{dB}$)}
    \label{fig:fig55}
\end{minipage} 
\end{figure}

In Fig. ~\ref{fig:fig5}, we plot the average power consumption and the loss function value of DREEM-NET as functions of the number of training episodes. In this test, we set $R_{\text{min}} = 0.2\,\text{bps/Hz}$ and SNR = $10\,\text{dB}$. We observe that both the average power consumption and the loss function decrease with the number of episodes, which demonstrates that the training process of DREEM-NET is carried out properly in a way of reducing the energy consumption in UDN.

In Fig. ~\ref{fig:fig55}, we plot the ratio of feasible solution in the action space as a function of mobile's rate requirement. We observe that when compared to the vanilla DQN scheme (basic DQN without special treatment), DSN in proposed DREEM-NET effectively reduces the action space. In particular, DREEM-NET achieves 58$\%$  reduction of the action space when $R_{\text{min}} = 0.5\,\text{bps/Hz}$, resulting in an improvement of cumulative reward and energy saving performance.


\begin{figure}[t]
\begin{minipage}[t]{0.48\linewidth}
   \includegraphics[width=\linewidth, height =70mm]{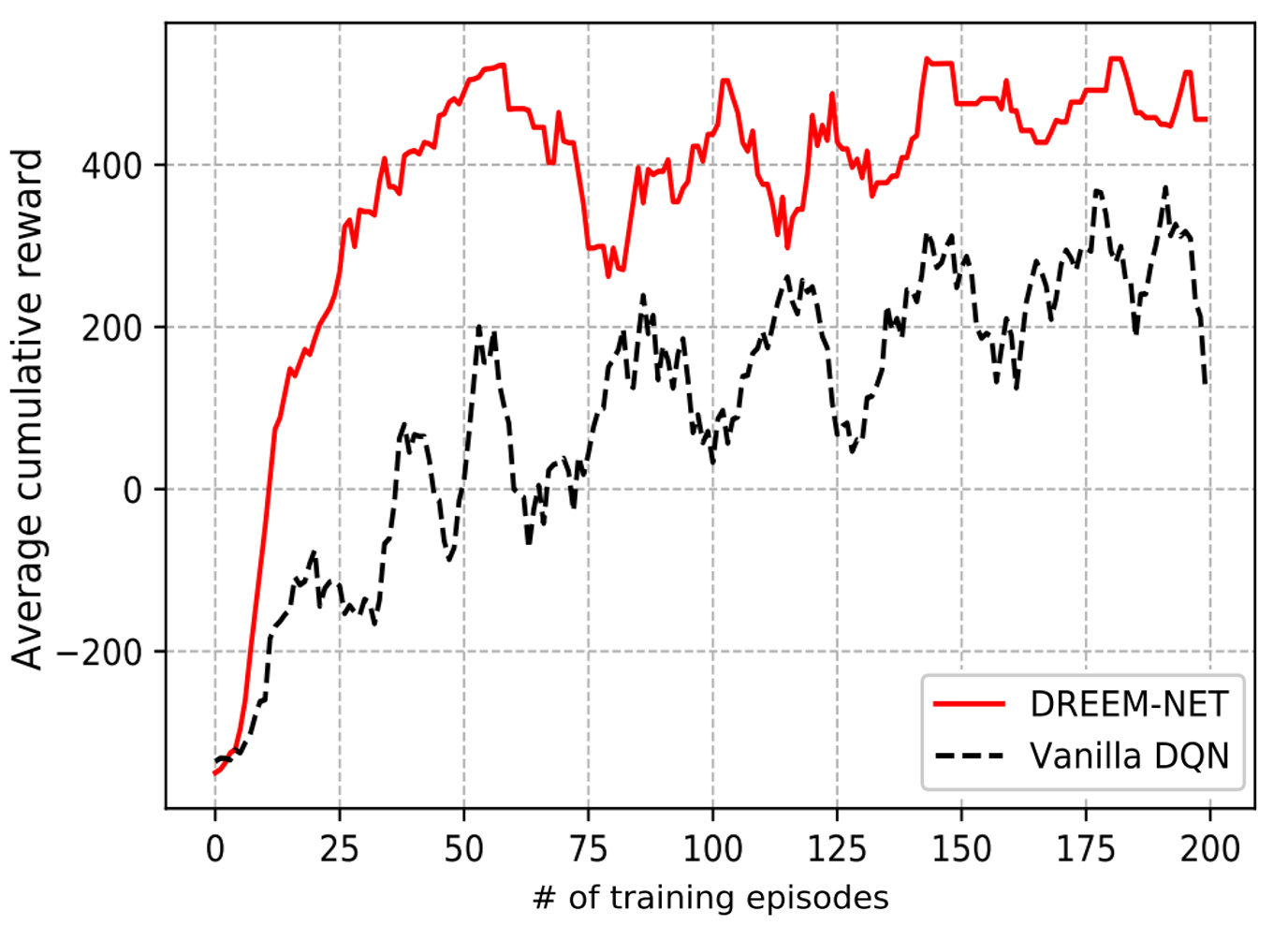}
    \caption{Average cumulative reward compared to vanilla DQN-based method in the training phase}
    \label{fig:fig11} 
\end{minipage}  
    \hfill%
\begin{minipage}[t]{0.48\linewidth}
    \includegraphics[width=\linewidth, height =70mm]{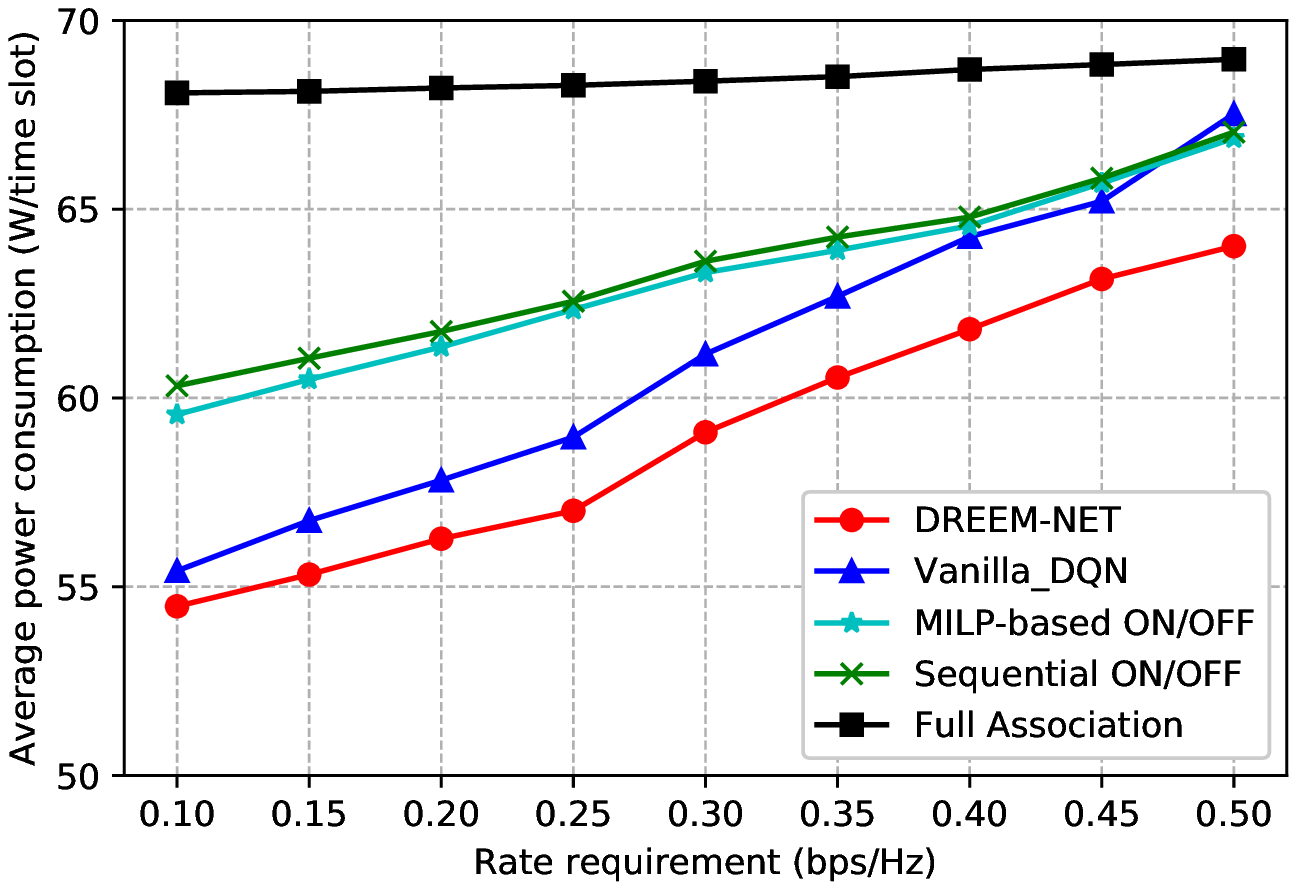}
    \caption{Average power consumption as a function of rate requirement $R_{\text{min}}$ ($M=10$, $K=4$, and $\text{SNR}=10\,\text{dB}$)}
    \label{fig:fig6}
\end{minipage}
\end{figure}

In Fig.~\ref{fig:fig11}, we plot the cumulative reward as a function of the number of training episodes in the training phase.
Since DREEM-NET receives fewer penalties by eliminating the infeasible decisions (see Sec.IV), the cumulative reward of DREEM-NET is much higher than that of the vanilla DQN-based method. 
Also, due to the elimination of the redundant decisions incurring excessive power consumption, DREEM-NET consumes less power than the vanilla DQN-based method.

In Fig.~\ref{fig:fig6}, we plot the average power consumption of the cell power control techniques as a function of mobile's rate requirement. We observe that the proposed DREEM-NET achieves considerable power saving over conventional methods.
For example, when $R_{\text{min}} = 0.1\,\text{bps/Hz}$, DREEM-NET saves more than 20$\%$ energy over the full association method and 12$\%$ over the MILP-based on/off algorithm.
This is because the MILP-based on/off scheme pursues a minimization of the instantaneous power (i.e., active/sleep mode power $P_m^{\text{mode}}$ and transmission power $P_m^{\text{tx}}$) while the proposed DREEM-NET controls the instantaneous power as well as the mode transition power $P_m^{\text{trans}}$.
Since the infeasible and redundant mode decisions are removed, DREEM-NET has better sample efficiency over the vanilla DQN-based method, resulting in an improvement in the energy saving ($5\% \sim 10\%$).

\begin{figure}[t]
\begin{minipage}[t]{0.48\linewidth}
    \includegraphics[width=\linewidth, height =70mm]{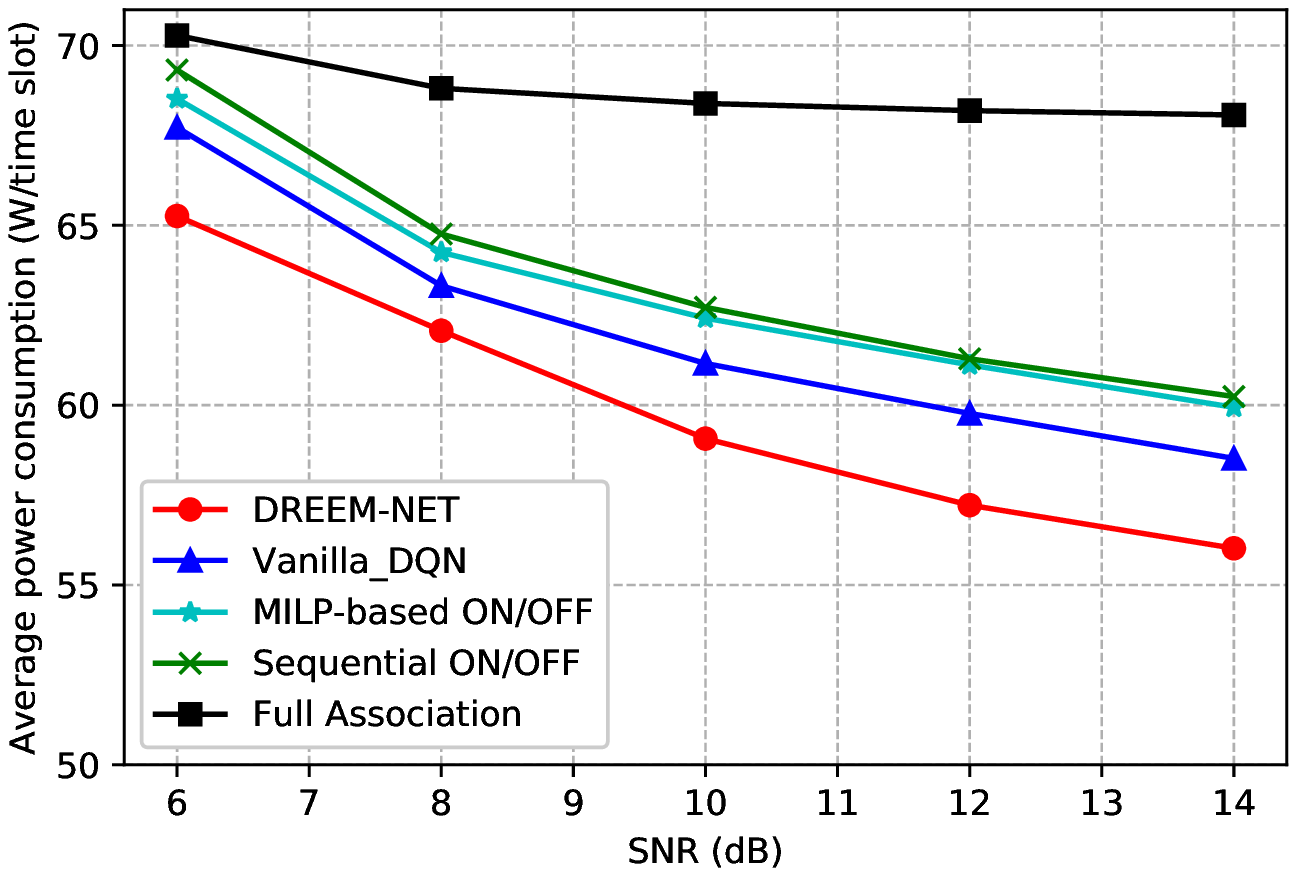}
    \caption{Average power consumption as a function of SNR ($M=10$, $K=4$, and $R_{\text{min}}=0.3\,\text{bps/Hz}$)}
    \label{fig:fig7}
\end{minipage}  
    \hfill%
\begin{minipage}[t]{0.48\linewidth}
    \includegraphics[width=\linewidth, height =70mm]{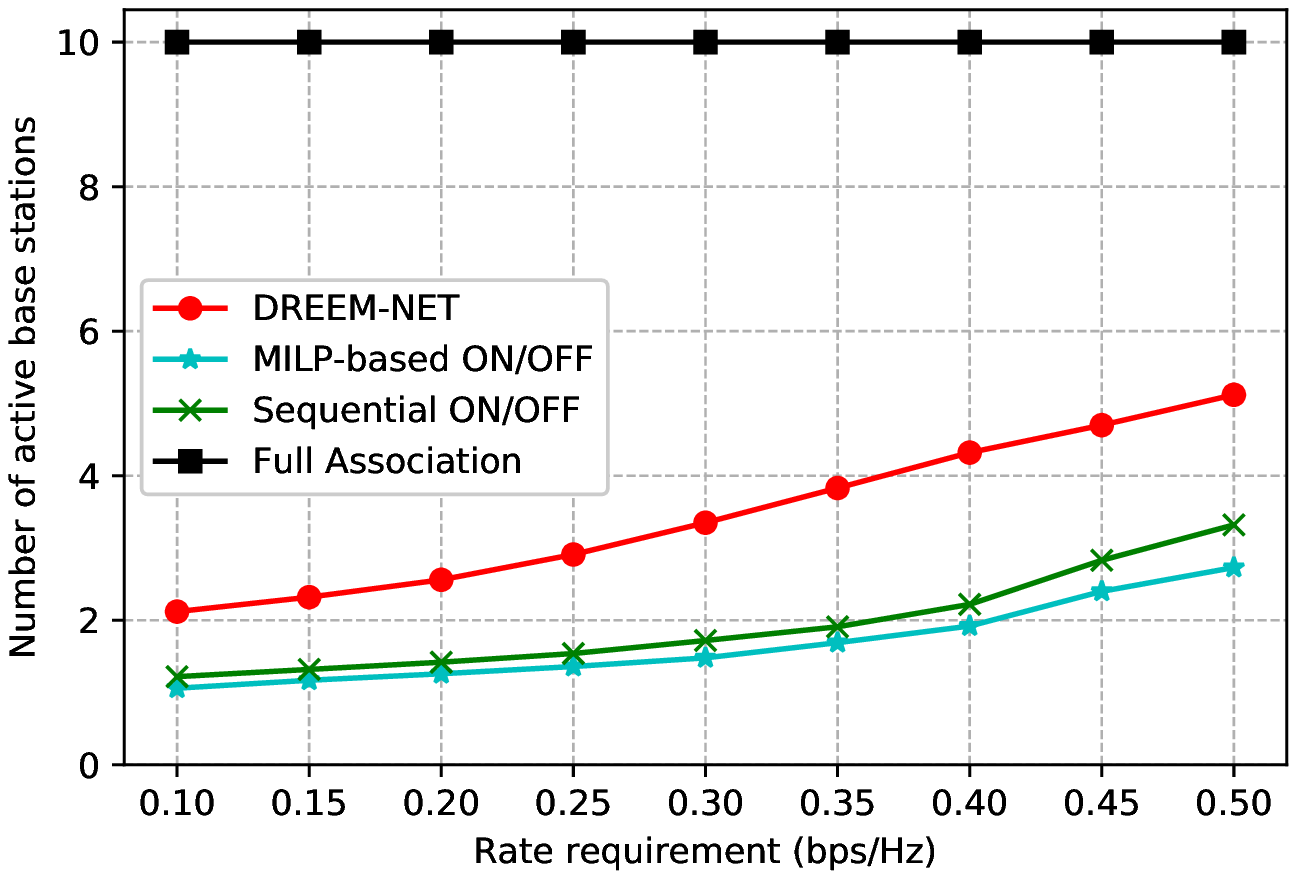}
    \caption{The number of active BSs as a function of rate requirement $R_{\text{min}}$ ($M=10$, $K=4$, and $\text{SNR}=10\,\text{dB}$)}
    \label{fig:fig8} 
\end{minipage} 
\end{figure}

In Fig.~\ref{fig:fig7}, we evaluate the average power consumption as a function of signal-to-noise ratio (SNR). 
We observe that the energy saving of the DREEM-NET scheme increases with SNR.
For example, when SNR = $6\,\text{dB}$, DREEM-NET saves about 6$\%$ power over the full association method but it goes up to 20$\%$ when SNR = $14\,\text{dB}$. 
This is because in high SNR, only a small number of active BSs is needed to serve the mobiles and thus substantial energy can be saved by turning off the lightly-loaded BSs.
We also observe that the energy saving of DREEM-NET over the conventional methods increases with SNR since the conventional methods typically choose active BSs near the mobile so that the active BS set is changed when the mobiles are moving. 
As a result, the energy saving of DREEM-NET over the MILP-based on/off method increases from 5$\%$ to 10$\%$ when SNR increases from $6\,\text{dB}$ to $14\,\text{dB}$. 
Even when compared to the vanilla DQN-based method, DREEM-NET saves around 6$\%$ energy on average.
Thus, the mode transitions of BSs occur frequently, causing a substantial increase in the mode transition power.


In Fig.~\ref{fig:fig8}, we plot the number of active BSs as a function of rate requirement when SNR = $10\,\text{dB}$. 
We observe that DREEM-NET turns on more BSs than the conventional on/off method does.
For example, when $R_{\text{min}} = 0.3\,\text{bps/Hz}$, DREEM-NET turns on $35\%$ of BSs whereas the sequential on/off method turns on only $20\%$ of BSs.
Nevertheless, DREEM-NET consumes 12$\%$ less energy than the MILP-based on/off method (see Fig.~\ref{fig:fig7}) because the saving of mode transition power outweighs the increase of maintenance power caused by turning on more BSs.

An important practical issue to be considered is the scalability of the system parameters such as the number of BSs or the number of mobiles.
If we need to re-train DREEM-NET whenever the system parameters are changing or the path loss model varies, computational complexity and training time in the training process will be unduly large, not to mention the large operating cost and effort.
Since the reconfiguration of BSs in UDN should be made in a few milliseconds (order of 1 ms subframe)~\cite{chen2016user}, frequent re-training is by no means suitable in the practical UDN scenario.
%

\begin{figure}[t]
\begin{minipage}[t]{0.48\linewidth}
    \includegraphics[width=\linewidth, height =70mm]{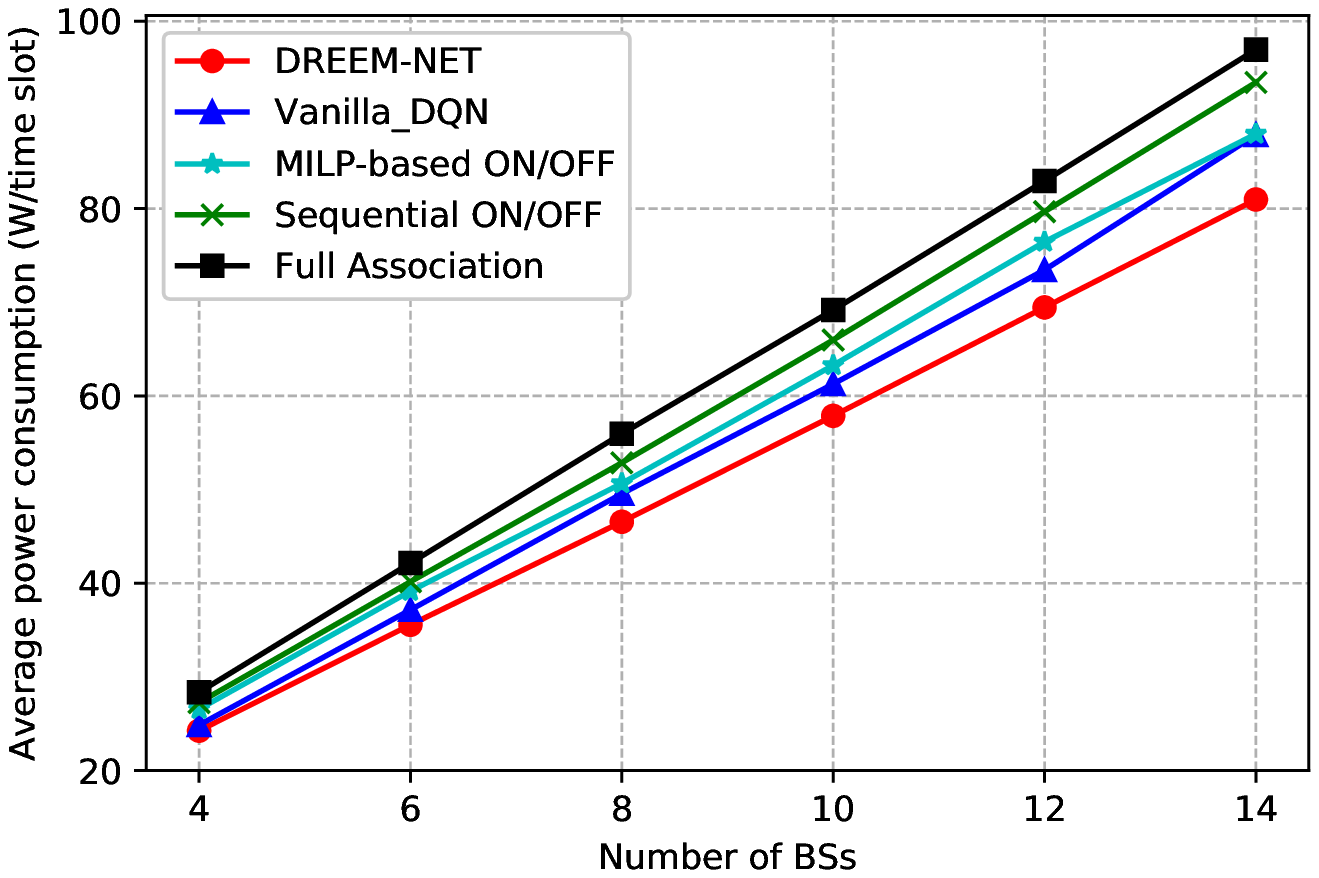}
    \caption{Average power consumption as a function of the number of BSs $M$ ($K=\frac{M}{2}$ and $R_{\text{min}}=0.2\,\text{bps/Hz}$)}
    \label{fig:fig9}
\end{minipage}  
    \hfill%
\begin{minipage}[t]{0.48\linewidth}
    \includegraphics[width=\linewidth, height =70mm]{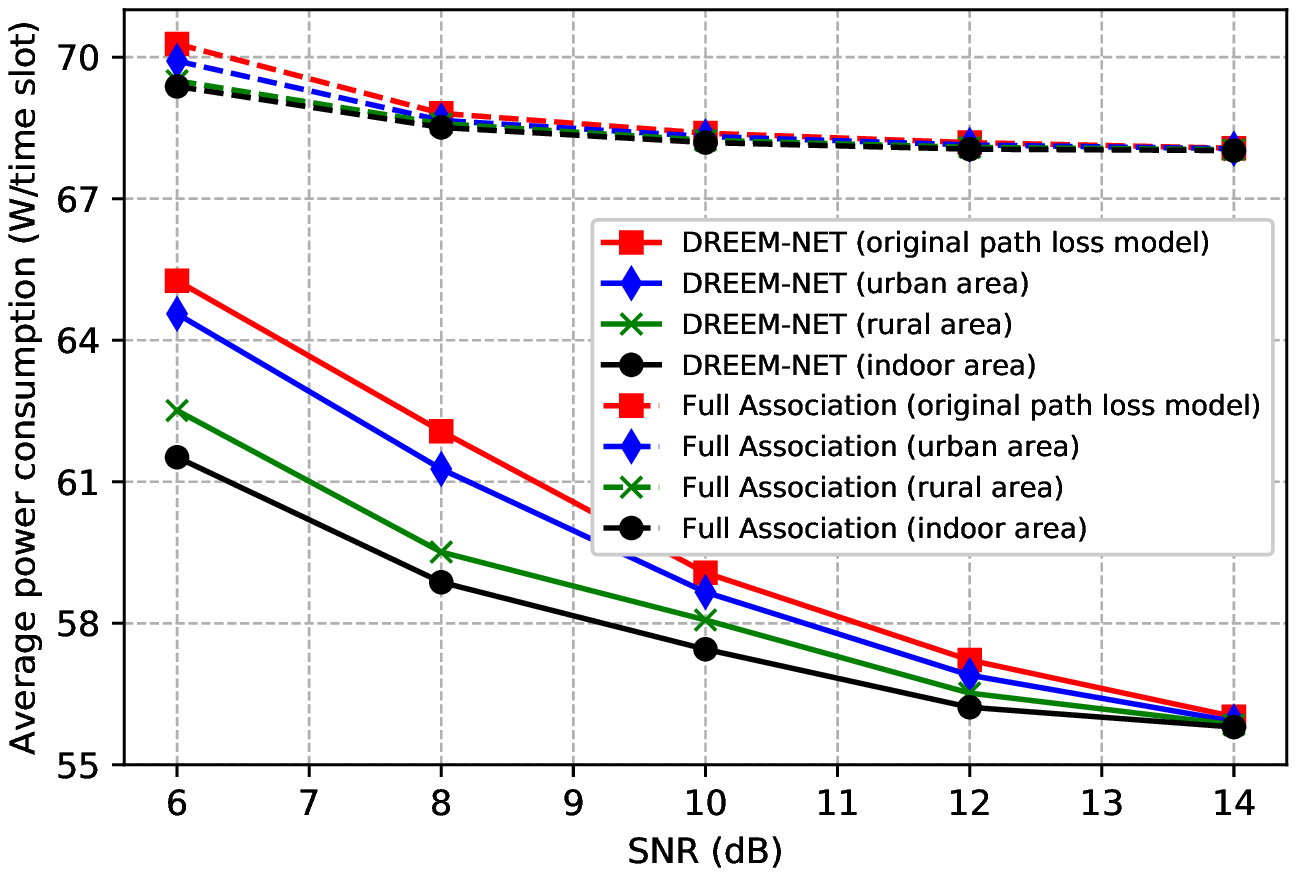}
    \caption{Average power consumption under various path loss models ($M=10$, $K=4$, and $\text{SNR}=10\,\text{dB}$)}
    \label{fig:fig10} 
\end{minipage}
\end{figure}

To investigate the scalability issue of DREEM-NET, we change the number of BSs and mobiles and then plot the average power consumption when $\text{SNR}=10\,\text{dB}$ and $R_{\text{min}}=0.2\,\text{bps/Hz}$.  
In this test, we train DREEM-NET for the ultra-dense scenario (i.e., $M=15$) and then test the scenarios with less number of BSs and mobiles. This is done by setting the input corresponding to the inactivated BSs and mobiles to zero and considering the inactivated BSs to be in sleep mode during the Q-value estimation process in DQN. 
We observe that the proposed training strategy works reasonably well with the marginal loss over the strategy that trains the DREEM-NET instances separately in each scenario. 
In particular, when $M=10$, the average power consumption of the proposed strategy is $57.6\,\text{W}/\text{slot}$ which is similar to the average power consumption of optimized DREEM-NET in Fig.~\ref{fig:fig6} ($57.1\,\text{W}/\text{slot}$). 
We also observe that DREEM-NET shows a considerable energy saving over the conventional on/off methods even in the ultra-dense scenario. Specifically, when $M=14$, DREEM-NET saves $10\%$ and $18\%$ of energy over the MILP-based on/off method and the full association method, respectively.

To test the robustness of DREEM-NET over the channel model change, we evaluate the performance of DREEM-NET under various path loss models (i.e., urban area, rural area, and indoor area) in 3GPP Release 14~\cite{zhu20193gpp}. 
Interestingly, we observe that DREEM-NET works well even when the tested scenario is different from the training scenario. 
Main reason for this is because the dynamic active/sleep mode decision of DREEM-NET relies heavily on the communication distance and the user's mobility rather than the specific channel model.

\begin{figure}[t]
\begin{minipage}[t]{0.48\linewidth}
    \includegraphics[width=\linewidth, height =70mm]{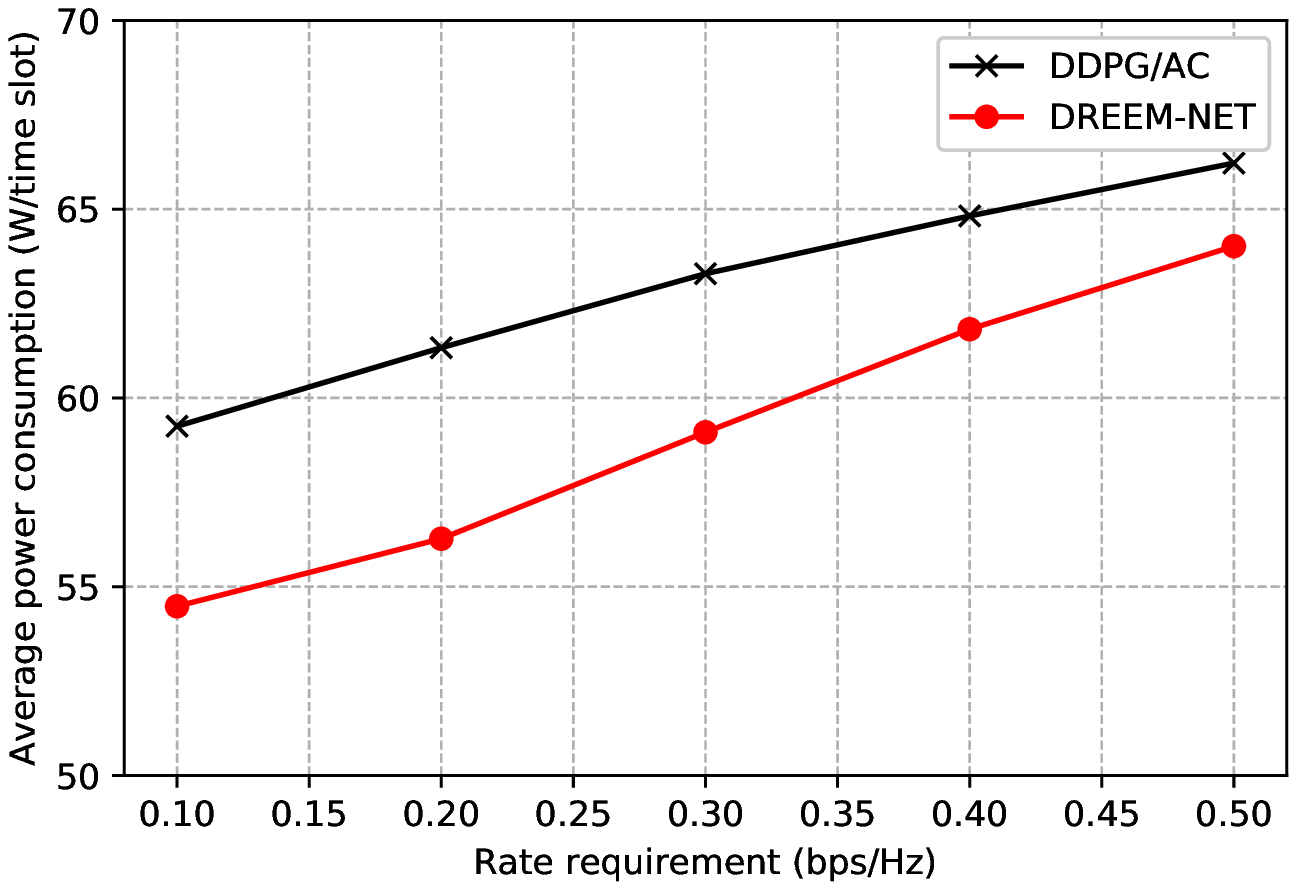}
    \caption{Average power consumption compared to DDPG/AC-based method ($M=10$, $K=4$, and $\text{SNR}=10\,\text{dB}$)}
    \label{fig:fig13}
\end{minipage}  
    \hfill%
\begin{minipage}[t]{0.48\linewidth}
    \includegraphics[width=\linewidth, height =70mm]{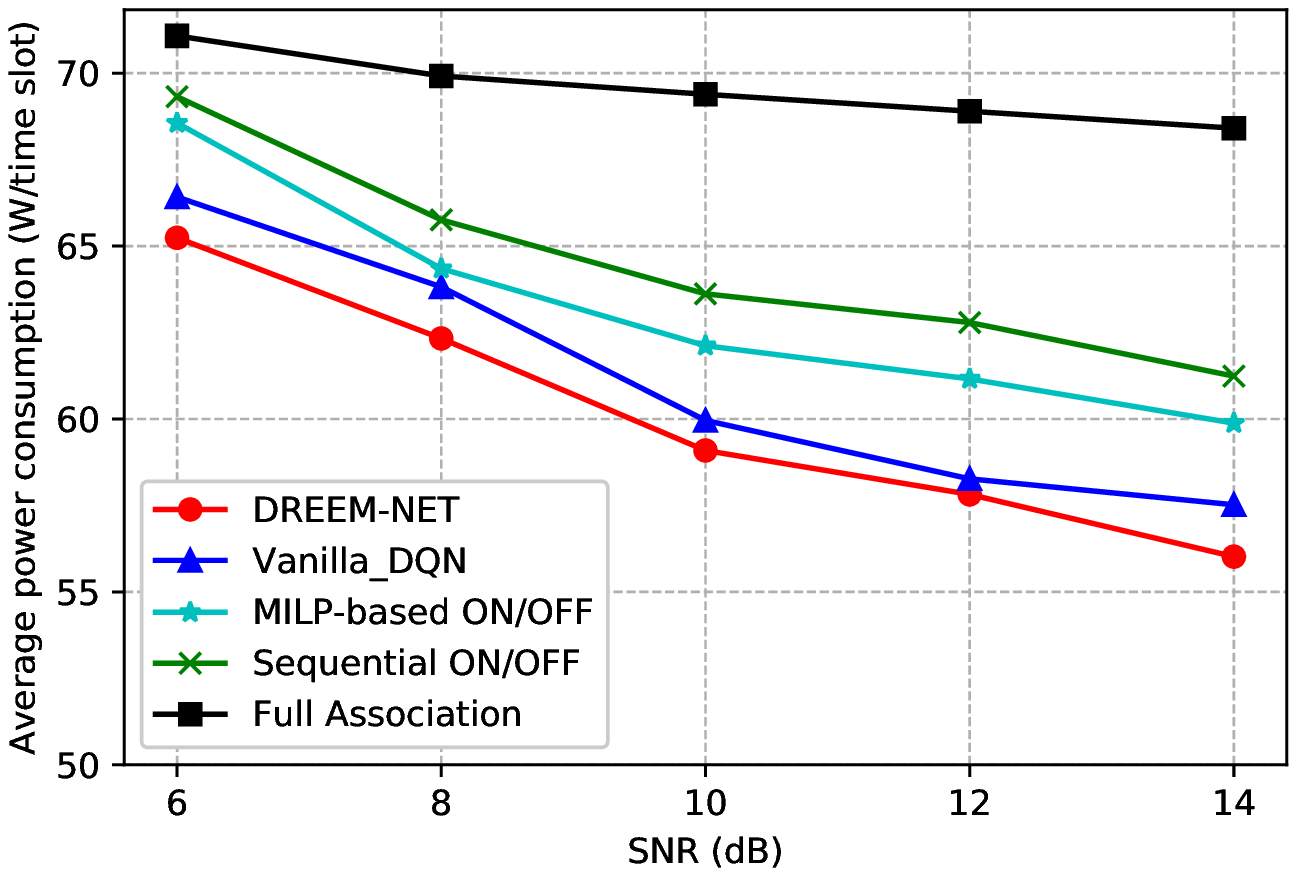}
    \caption{Average power consumption under the time-varying traffic model ($M=10$, $K=12$)}
    \label{fig:fig14} 
\end{minipage}
\end{figure}

In Fig.~\ref{fig:fig13}, we compare the performance of proposed scheme with the DDPG/AC-based active/sleep strategy. In general, since the DRL agent learns the decision-making policy by trial and error, the performance of DRL technique depends heavily on the exploration of action space (i.e., sample efficiency).
The reason that the conventional DRL-based schemes (e.g., DDPG and vanilla DQN) do not perform well is because the DRL agent is likely to explore undesirable actions violating the mobile's rate requirement or turning on too many BSs due to the humongous discrete action space.
In contrast, in the proposed DREEM-NET, by eliminating the undesirable actions via feasibility test and energy consumption test, the chance of exploring desirable actions increases considerably.
Also, we would like to point out that the DDPG/AC technique can effectively handle the continuous action space by exploiting the property that Q-values of adjacent continuous actions are fairly similar. However, in the discrete action space, Q-values of adjacent actions may differ greatly and thus, the property of the continuous action space is not that useful.

In Fig.~\ref{fig:fig14}, we plot the average power consumption under the time-varying traffic model (i.e., FTP model 3). We observe that DREEM-NET outperforms the conventional methods under the time-varying traffic scenario.
 
\section{Conclusion}
In this paper, we proposed the DRL-based BS sleep mode decision framework, referred to as DREEM-NET, to improve the energy efficiency in UDN. 
In the proposed DREEM-NET, the infeasible or redundant active/sleep mode decisions are eliminated by specially designed decision selection network.
In doing so, we could reduce the active/sleep mode decision space significantly, thereby achieving an improvement of the sample efficiency and energy saving.  
Further, the proposed DREEM-NET saves not only the instantaneous power but also the mode transition power so that we could reduce the cumulative energy consumption.
From the numerical evaluation, we demonstrated that DREEN-NET saves up to $20\%$ of energy consumption against the full association scenario and the vanilla DQN-based method. In this paper, we restricted our attention to the energy efficiency improvement of UDN but we expect that the proposed scheme can be extended to various tasks such as cognitive radio access, user scheduling, and resource allocation.
\\
\bibliography{refs_journal}
\bibliographystyle{IEEEtran}

\end{document}